\documentclass[3p]{elsarticle}

\usepackage{lineno,hyperref}
\usepackage{xcolor}
\usepackage{physics}
\usepackage{txfonts}
\usepackage{multirow}
\usepackage{xspace}
\usepackage{graphicx}
\usepackage{setspace}
\usepackage{subcaption}  % for subfigure layout

\newcommand{\mycomment}[1]{}
\modulolinenumbers[5]

\begin{document}

\begin{frontmatter}

\title{ Material Identification using Multi-Modal Intrinsic Radiation and Radiography\tnoteref{mytitlenote}}
%\tnotetext[mytitlenote]{Fully documented templates are available in the elsarticle package on \href{http://www.ctan.org/tex-archive/macros/latex/contrib/elsarticle}{CTAN}.}

%% Group authors per affiliation:
\author[lanl]{Khoa Nguyen\corref{mycorrespondingauthor}}
\cortext[mycorrespondingauthor]{Corresponding Author}
\ead{<nguyen_k@lanl.gov>}
\author[lanl]{Brendt Wohlberg}
\ead{<brendt@lanl.gov>}
\author[lanl]{Oleg Korobkin}
\author[lanl]{Marc Klasky}

\address[lanl]{Los Alamos National Laboratory, Los Alamos, NM 87545, USA}
\ead{<mklasky@lanl.gov>}

%\fntext[myfootnote]{Since 1880.}

\begin{abstract}
    We investigate multi-modal material identification for special nuclear material (SNM) configurations using a combination of X-ray radiography, high-resolution $\mathrm{\gammaup}$-ray spectroscopy, and neutron multiplicity measurements. We consider a Beryllium Reflected Plutonium sphere  (BeRP) ball surrounded by one or two concentric shielding shells of unknown composition whose radii are assumed known from radiography. High-purity germanium (HPGe) spectra are reduced to net counts in selected Pu-239 photo-peaks, while neutron multiplicity information is summarized by Feynman variances $Y_2$ and $Y_3$ computed from factorial moments of the neutron counting statistics. Using synthetic data generated with the Gamma Detector Response and Analysis Software (GADRAS) for a range of shielding materials and thicknesses, we cast the material identification problem as a supervised multi-class classification task over all admissible shell-material combinations. We demonstrate that a random forest classifier trained on combined gamma and neutron features achieves almost perfect identification accuracy for single-shell cases, and substantial performance gains for more challenging double-shell configurations relative to gamma-only classification. Alternative statistical and machine-learning formulations for this multi-class problem are examined along with examination of the impact of model-mismatch between the forward model and the test cases as given by variations in the statistical noise. Opportunities for extending the approach to more complex geometries and experimental data are also discussed.  

   % In this work we investigated multi-modal material identification for spherically symmetric shielding configurations using a combination of X-ray radiography, gamma-ray spectroscopy, and neutron multiplicity measurements. For single-shell configurations with known shell thickness, gamma spectroscopy alone provides excellent material discrimination for moderate and large thicknesses, while thinner shells (0.5~in) exhibit residual confusion among materials with similar attenuation properties. The inclusion of neutron multiplicity features, particularly the second-order Feynman variance $Y_2$, consistently reduces these confusions by introducing sensitivity to moderation and absorption properties that are complementary to gamma attenuation. For double-shell configurations, the classification task becomes substantially more challenging due to the combinatorial increase in material configurations and ordering effects. In this regime, gamma-only performance degrades significantly and exhibits pronounced asymmetry with respect to shell ordering, whereas the incorporation of neutron multiplicity measurements yields large performance gains and substantially mitigates ordering-dependent ambiguities. Although residual asymmetries remain due to the inherently directional nature of gamma-ray transport in layered media, the results demonstrate that multi-modal gamma–neutron measurements pro_

\end{abstract}


\end{frontmatter}

% \linenumbers

\section{Introduction}
\label{sec:intro}

The use of intrinsic radiation to perform material identification is valuable in a variety of medical, baggage screening, geophysics, industrial, nonproliferation and nuclear security applications~\cite{alvarez1976, kalender1986, vetter1986, yingz2006, fraser1986, paziresh2016, cann1982}. While advances in material identification in the medical and baggage scanning applications have been realized using dual energy tomography, additional challenges have limited advances in the industrial and nuclear material accountability and security settings\textcolor{red}{~\cite{alvarez1976,runkle2009}}.  Among the major challenges in addressing material identification in the nuclear security and industrial settings is the necessity to examine thick objects, i.e., those with large areal mass. This regime presents challenges, even for a low effective nuclear charge ($Z_{\rm eff}$) material composition, in so far as the X-rays that are typically used in medical and baggage-screening applications have energy ranges (70–100 and 135–150 kVp) that are too low to penetrate the object.   Attempts have been made to address these additional challenges by employing more penetrating radiation, i.e. those with MeV scale~\cite{ogorodnikov2002, naydenov2004}. 
However, imaging in this energy regime suffers from minimal contribution from the photoelectric effect, whose $Z^4$ scaling is largely responsible for the ability to perform material identification. 
Moreover, in this energy regime, attenuation is nearly constant across both the Compton window and the onset of pair production. Attempts to perform material identification exploiting the $Z^2$-dependence at high energy, i.e. 10 MeV, is possible, but these sources are generally not available in the industrial and security arenas, and the overlap in spectra using these poly-energetic spectra is significant.
Finally, in both industrial and nuclear security applications, the X-rays may traverse multiple materials, further limiting the application of traditional dual energy CT algorithms, only providing an average $Z_{\rm eff}$ and not isolation of the material composition of the constituents~\cite{tobias2019, runkle2009}.  

The use of counting detectors has opened a potentially promising route to enhance the ability to perform material identification in nuclear security and industrial settings ~\cite{beldjoudi2012,wux2017}. It has also been recognized that monoenergetic sources offer greater enhancement in performance relative to polychromatic sources \cite{mendoncca2013, xuey2019, mendoncca2010, long2014, kelcz1979}. However, these high-energy monoenergetic sources are not readily available ~\cite{runkle2009}. If prior information regarding the interfaces of materials can be obtained via segmentation of a 3D object, a discrete material identification problem may be formulated ~\cite{herman2007}. Indeed, a material identification algorithm based on using a radiograph and a single integrating flat-panel detector was demonstrated to enable material identification of concentric shells using both synthetic as well as experimental data~\cite{mccann2023}.  Finally, an optimization methodology was developed to perform material identification using radiography in conjunction with line emissions from SNM in conjunction with a HPGe counting detector to further enhance the ability to perform material identification.\cite{korobkin2024isotopic} While this approach was successful in many cases, it exhibited significant difficulty in performing material identification in others.

 The use of neutron measurements for material identification has recently been examined. These investigations have  focused on hyperspectral measurements which typically require extremely long count times, making them impractical for real time applications. \cite{tang2024machine, chowdhury2023autonomous}  An alternative is to employ the neutron multiplicities to enable material identification while diminishing the count time. It should be noted that while extensive work has been devoted to the study of neutron multiplicity, it has been almost exclusively used for estimates of Pu mass for nuclear control and accountability.\cite{ensslin1991principles, hage1985factorial, dubi2018mass, cifarelli1986models}. Furthermore, although there have been suggestions that neutron multiplicity measurements may be used to infer material composition of accompanying materials surrounding SNM, we are unaware of any comprehensive work in this area. Consequently, we have performed preliminary material identification calculations using neutron multiplicity measurements to enable material identification. \cite{LA-UR-24-22644,LA-UR-23-23930,LA-UR-24-22707}

\begin{figure*}[t]
    \centering
    
    %-------------------- Row 1: shell thickness 0.5 in --------------------%
    \resizebox{1.0\columnwidth}{!}{%
    \begin{subfigure}[t]{\columnwidth}
        \centering
        \includegraphics[width=\linewidth]{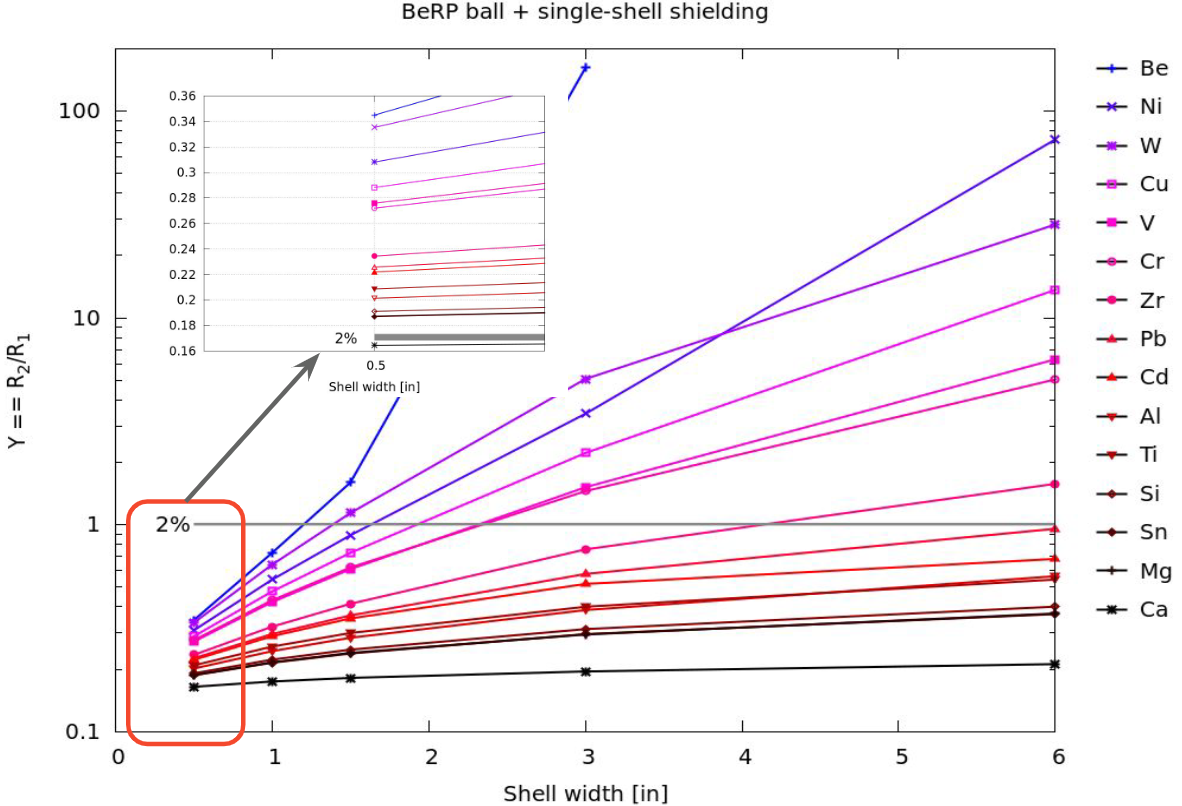}
        \caption{$Y_2$}
        \label{fig:Y2_pt5}
    \end{subfigure}
    \hfill
    \begin{subfigure}[t]{\columnwidth}
        \centering
        \includegraphics[width=0.9\linewidth]{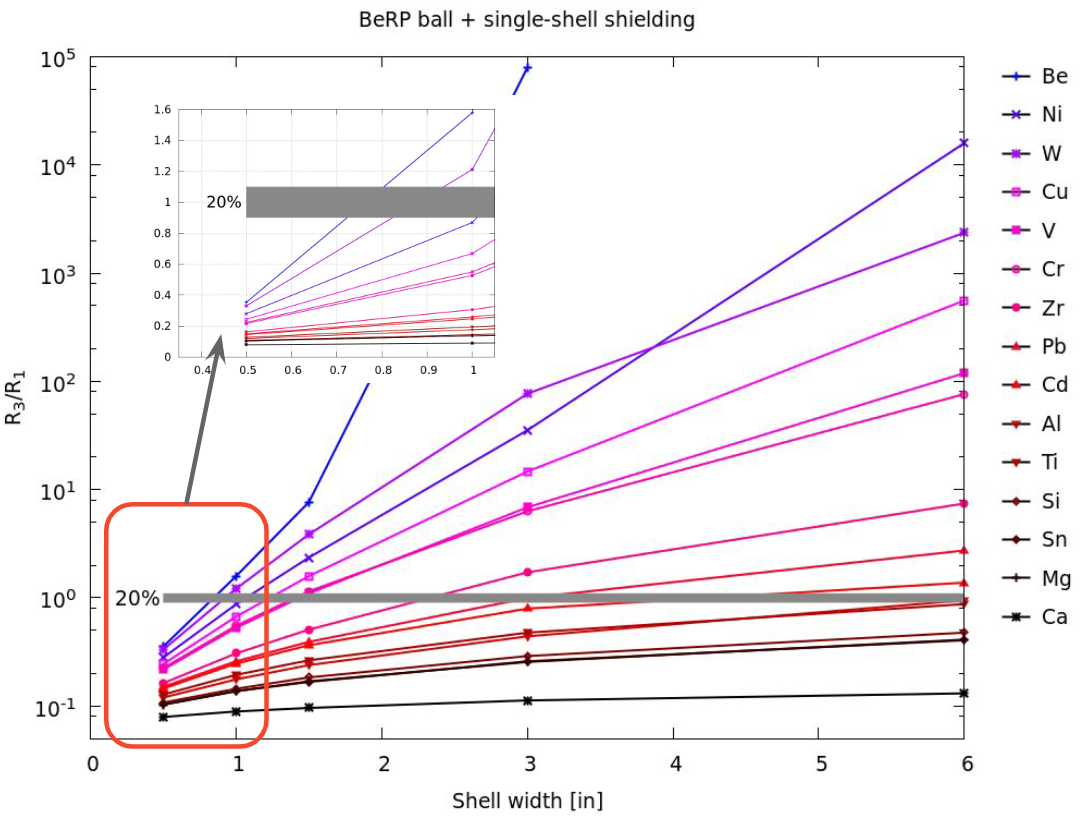}
        \caption{$Y_3$}
        \label{fig:Y3_pt5}
    \end{subfigure}
    }
    \caption{Feynman Variance {$Y_2$} and {$Y_3$} for BeRP ball versus shielding shell thickness }
\end{figure*}

%We do remark however, that the use of hyperspectral neutron imaging has been utilized to determine material composition.\cite{tang2024machine,chowdhury2023autonomous}  
%show the feasability of the approach
  As an illustration of the computational problem setting, we present an examination in which a BeRP ball is shielded by a shell of unknown composition whose outer radius has been determined from a radiographic image using a traditional edge finding algorithm.  Using the Gamma Detector Response and Analysis Software (GADRAS), we computed the neutron multiplicity moments using an M15 detector at a radius of 1 meter from the outer shell surface~\cite{mitchell2009}.  Results of this examination are presented in Figures~\ref{fig:Y2_pt5} and ~\ref{fig:Y3_pt5}.
  
Examination of Figures~\ref{fig:Y2_pt5} and~\ref{fig:Y3_pt5} reveals that for fixed Pu composition, mass, and shell thickness, the material composition of the shielding may indeed be determined via the Feynman variances Y2 and Y3.  That is, for a known shell thickness, and isotopic composition of the source, the Y2 and Y3 responses are generally different for most materials.  However, it is noted that the ability to differentiate materials will be impacted by the uncertainty in the multiplicity measurement.  Experimental evidence suggest that the error in the Y2 measurement is approximately two-percent{~\cite{LAPR-2021-037525}.  In addition, higher order factorial moments of the neutron distribution may also be utilized to provide additional discrimination ability.  Here we utilize Y3 in our investigations and utilize an uncertainty of twenty percent~\cite{LAPR-2021-037525}.  Additional discussion on the uncertainty in these measurements as well as a parametric investigation on its impact on material identification are presented in Sections 3 and 5.2.

While material identification has been investigated using  traditional radiography, gamma spectroscopy, as well as neutron multiplicity, to our knowledge material identification using multiple modalities has not previously been performed. In this work, we report on a multi-modal technique using intrinsic radiation (neutron multiplicity measurements as well as gamma radiation) in conjunction with x-ray radiography to improve the ability to perform material identification.   In Section 2 we present our problem setting and data generation methodology.  This specific application is one in which each segmented region of a radiographic image contains only a single material with a known density and associated attenuation coefficient. An illustration of one such object is depicted in Figure~\ref{fig:problem-formulation} \cite{korobkin2024isotopic}. We demonstrate our proposed algorithm using simulations on spherically symmetric geometries with multiple layers of shielding materials. 
%In this work we also demonstrate the improvement in determining the amount of mass of nuclear material when the material composition of the accompanying concentric shells is properly identified. 
In Section 3 we describe our data pre-processing for both gamma spectroscopy and neutron multiplicity. Section 4, presents the problem formulation for multi-modal material identification using both neutron multiplicity as well as gamma spectroscopy in conjunction with radiography to determine the material interfaces.   Section 5 presents the results of our investigations.   Section 6 provides discussion and conclusions.  Finally, Section 7 provides suggestions for future investigations.

Our specific contributions are
\begin{enumerate}
\item The formulation of a rapid multi-modality material identification algorithm;
\item Demonstrations of the improvements realized by the inclusion of neutron multiplicity in material identification relative to using gamma spectroscopy alone. 
\item Examination of the ability of different ML algorithms to perform material identification.
\end{enumerate}

\section{Problem Formulation and Data Generation}
\label{sec:problem}

The general problem that we examine is one in which a collection of objects are hidden within a scene. 
Their positions, orientations, and most importantly, materials of potential concern, are interrogated with penetrating radiation (see Fig.~\ref{fig:problem-formulation}).

\begin{figure*}[t]
    \centering
    \begin{tabular}{p{0.32\textwidth}p{0.20\textwidth}p{0.35\textwidth}}
    \includegraphics[width=0.32\textwidth]{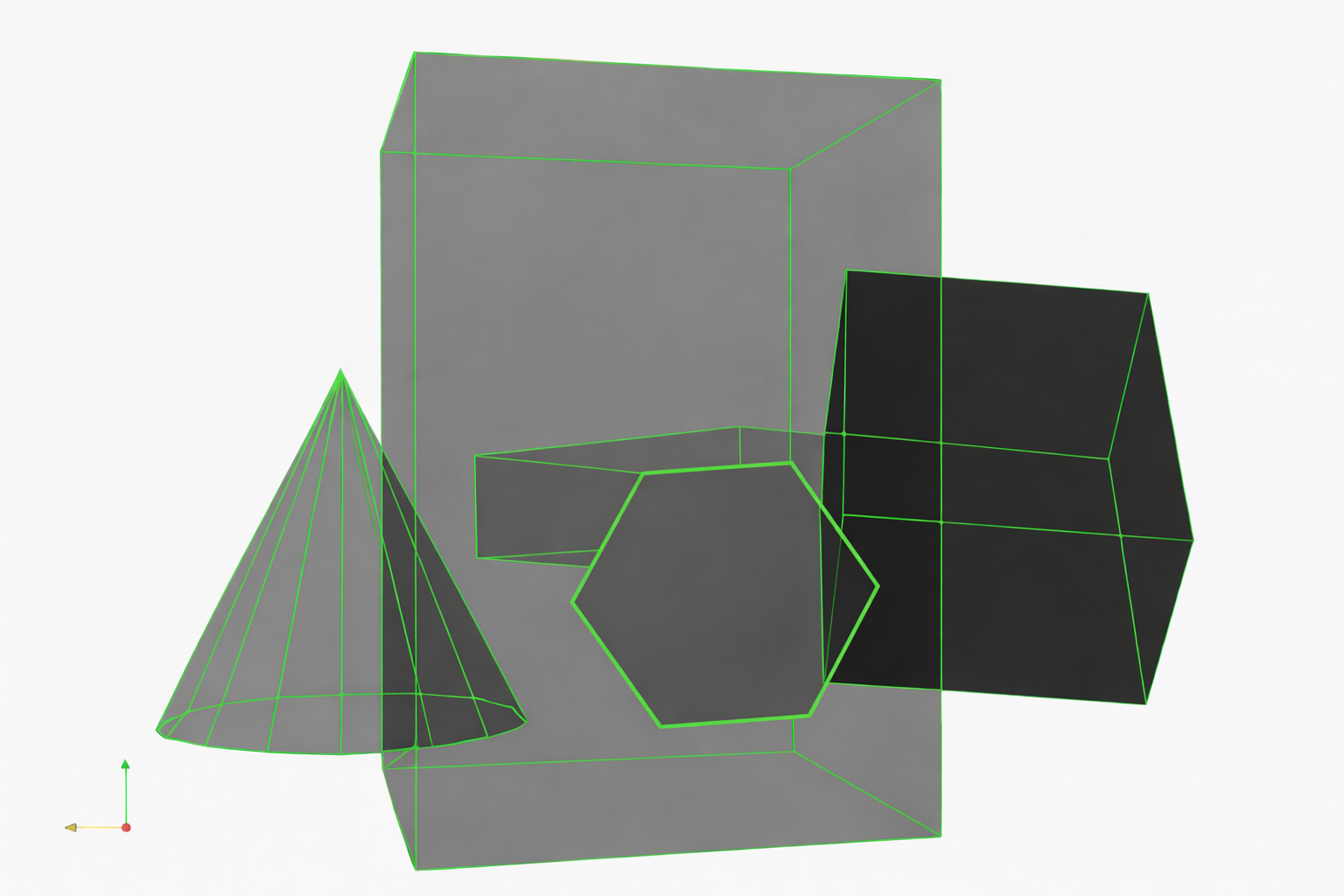} &
    \includegraphics[width=0.20\textwidth]{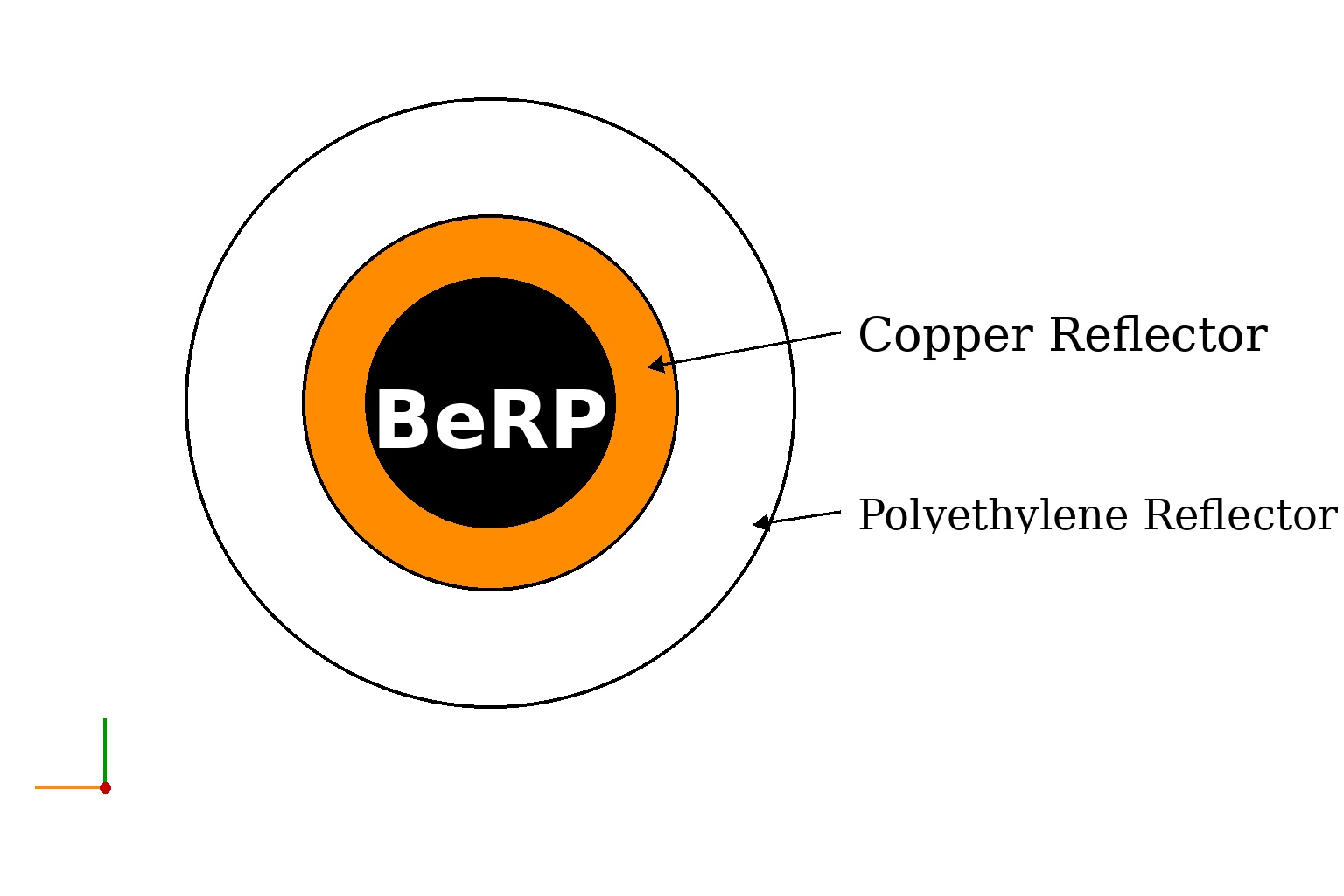} &  
    \includegraphics[width=0.3\textwidth]{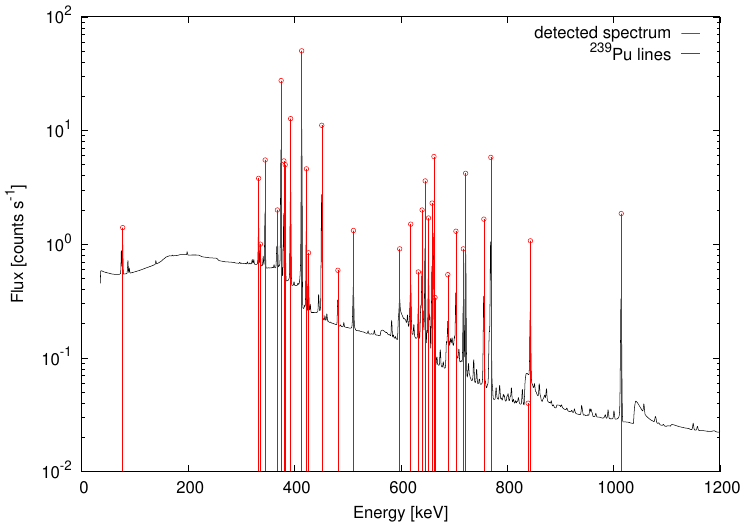}
    \\
    {\footnotesize (a) Segmentation of a synthetic image} &
    {\footnotesize (b) Static BeRP Ball  with two shielding layers. } &
    {\footnotesize (c) High-resolution spectrum and lines.}
    \end{tabular}
    \caption{General problem formulation.}
    \label{fig:problem-formulation}
\end{figure*}
As an initial step, objects are scanned with X-rays, revealing their positions and geometries, possibly with contour finding algorithms obtained from radiographs, from various angles if needed.
We assume here that the geometry or the scene can be recovered with sufficient accuracy using tomographic and segmentation algorithms. Indeed, we have demonstrated that recovery of complex topologies with very limited views is indeed possible using sparse view tomographic algorithms~\cite{xu2025swap,lahiri2023sparse}.
In the problem we are considering, one of the objects contains a radioactive gamma-ray emitting material, such as $^{235}$U or $^{239}$Pu, possibly surrounded by various shielding shells.
Assuming that the presence of this gamma-emitting isotope has been determined~\cite{khatiwada23},
we seek to identify the shielding materials by acquiring a highly resolved energy spectrum of the intrinsic radiation.  For our purposes, only the presence of a certain gamma-emitting isotope needs to be established; knowledge of its exact amount is not required.  
A spectrum with energy resolution necessary to resolve individual lines ($\approx1\%$) can be taken using e.g. high-purity germanium (HPGe) detectors~\citep{knoll2010}. 

The workflow is illustrated in Fig.~\ref{fig:problem-formulation}, where a segementation of sparse view tomographic reconstruction of a hypothetical synthetic scene, has revealed a scene in which material identification is desired. \cite{xu2025swap,bell2025learning}.
Figure~\ref{fig:problem-formulation}b shows an example composition of this spherical object, with the innermost shell containing $^{239}$Pu isotope.
Gamma radiation emitted by this radioactive source is attenuated by shielding materials, and the simulated spectrum is shown with a black solid line in Fig.~\ref{fig:problem-formulation}c.
Red lines correspond to the net detector events under each photopeak of $^{239}$Pu to be used for identification of the attenuating materials.  Neutron multiplicity measurements are recorded by a neutron multiplicity detector, from which the Feynman variances may be computed.

In our investigations we consider a 4.5 kg sphere of $\alpha$-phase weapons-grade
plutonium metal, originally manufactured in October 1980 by Los
Alamos National Laboratory, commonly referred to as the BeRP ball~\cite{mattingly2009polyethylene}. This sphere has a mean radius of
3.7938 cm, and is encased in a 304 stainless steel shell that is
0.0305 cm thick. The plutonium is composed of $\alpha$-Pu with an isotopic composition as presented by Mattingly~\cite{mattingly2009polyethylene}.

We perform our investigations on the ability to perform material identification using the BeRP ball with both one and two accompanying shielding shells of varying thickness, specifically 0.5, 1.0,1.5, 3, and 6 inches. For these configurations the following materials were utilized.

%\begin{figure*}[!htbp]
    
    %\begin{tabular}{cc}
      %\includegraphics[width=0.55\textwidth]{single shell(1).pdf} &
      %\includegraphics[width=0.55\textwidth]{double_shell.jpg}
      % \\
      
%    \end{tabular}
    %\caption{Representative Single and Double BeRP Ball Shielding Configurations .} % caption
    %\label{fig:id_single_double_mat}
%\end{figure*}

\begin{itemize}
    \item 1-shell materials: 'Al', 'B', 'Be', 'C', 'Ca', 'Cd', 'Cr', 'Cu', 'Fe', 'HDPE', 'Mg', 'Ni', 'Pb', 'Si', 'Sn', 'Ta', 'Ti', 'V', 'W', 'Zn', 'Zr'
    \item 2-shell materials: 'Al', 'B', 'Be', 'C', 'Ca', 'Cu', 'Fe', 'HDPE', 'Ni', 'Pb', 'Ta', 'Ti', 'W', 'Zn'    
\end{itemize}
%  Representative examples for the two configurations are presented in Fig.~\ref{fig:id_single_double_mat}.  

For each of these respective configurations, gamma spectra as well as neutron multiplicity signals were  generated using  GADRAS \cite{mitchell2009}. A LANL~S HPGe detector was utilized for the gamma spectroscopic signal and a MC-15 detector for the neutron multiplicity. Finally, the detectors were located approximately 1~m from the objects, with an  elevation of 30 cm, and the collection time in all experiments was 300~s. 

\section{Data preprocessing and feature extraction}
\label{sec:dataprep}

GADRAS simulations for the gamma spectroscopic measurements produce a continuous spectra, as shown in Figure~\ref{fig:problem-formulation}c  black solid line with units of counts per channel. This continuous curve may contain multiple contaminants irrelevant to the problem, including environmental background, scattering, finite line width, etc in additional to the Pu-239 emission lines. In these investigations we utilize the previously determined Pu-239 spectroscopic lines to facilitate the best material identification via gamma spectroscopy~\cite{korobkin2024isotopic}.

%It should be noted that in these investigations we do not perform pre-processing to remove the continuum but rather utilize the noise-realizations as was previously identified. \cite{khatiwada2023machine}

%To facilitate our multi-modality material identification algorithm we perform continuum subtraction and obtain the net-counts for each photo-peak of interest by integrating around the photo-peaks using the PEAKEASY software.\cite{rooney2018peakeasy}

In the present work we utilize total photopeak counts without explicit continuum subtraction. This choice is motivated by previous investigations into machine-learning–based isotopic determination using HPGe $\gamma$-ray spectra, which demonstrated that explicit continuum subtraction provides minimal benefit and can, in some cases, slightly degrade performance due to reduced counting statistics and additional preprocessing uncertainty~\cite{khatiwada2023machine}. In that study, comparable or improved accuracy was obtained across multiple ML algorithms for both shielded and unshielded configurations when continuum subtraction was omitted, indicating that the algorithms were able to learn discriminative feature relationships directly from spectra containing substantial continuum contributions. Given that the present work addresses a classification problem using ensemble methods with noise augmentation, the use of total photopeak counts is both justified and consistent with established results.

The spectral lines are as follows:
\begin{itemize}
     
    \item  (keV) 129.27, 203.53, 332.82, 344.99, 375.03, 392.91, 413.69, 451.46, 639.94, 645.87, 769.24
\end{itemize}

In addition to the gamma spectroscopic features, we utilize the neutrons emitted to enable additional information to be utilized to enhance the ability to perform material identification of the accompanying shielding materials.  To accomplish this we exploit the distributions of the emitted neutrons from the objects. These neutron may be characterized by neutron multiplicity moments that have been previously employed in the Nuclear Safeguards arena using neutron multiplicity detectors to quantify the amount and isotopic composition of fissile material (such as plutonium or highly enriched uranium) in various forms and containers, particularly when other measurement techniques (like gamma spectroscopy) are insufficient due to shielding or impurities~\cite{langner1998application}. Indeed, by using a neutron multiplicity detector we can obtain characteristics of the neutron distribution emissions i.e. singles, doublets, triplets etc.  These distributions can then be utilized to calculate the factorial moments, or reduced factorial moments, of the recorded neutron population from which the Feynman variances may be calculated~\cite{dubi2017estimating}.

The \textbf{factorial moment of order} $q$ is defined as:
\begin{align}
\overline{n(n-1)\cdots(n - q + 1)} 
&= \left. \frac{\partial^q G(z)}{\partial z^q} \right|_{z=1} \nonumber \\
&= M_q G \nonumber \\
&= \sum_{n=q}^{\infty} n(n-1)\cdots(n - q + 1) \, p_n \;.
\end{align}
Here, $M_q$ is the differential operator:
\begin{equation}
M_q = \frac{\partial^q}{\partial z^q} \;.
\end{equation}
The first three moments of the neutron multiplicity counting distribution $R_q$ are related to the factorial moments as follows:
\begin{align}
\overline{n} &= M_1 G = R_1 \\
\overline{n(n-1)} &= M_2 G = R_2 + R_1^2 \\
\overline{n(n-1)(n-2)} &= M_3 G = R_3 + 3R_1 R_2 + R_1^3 \;,
\end{align}
and the \textbf{``excess''} $Y_q$ for each factorial moment is defined by:
\begin{align}
Y_1 &= 0 \\
Y_2 &= R_2 \\
Y_3 &= R_3 + 3R_1 R_2 \;.
\end{align}
In our numerical experiments we use $Y_2$ and $Y_3$ as neutron features. The corresponding factorial moments and Feynman variances for each configuration are evaluated using GADRAS~\cite{mitchell2014gadras}.
% ===================== PATCH BEGIN =====================
\paragraph{Remark on source isotopics and neutron multiplicity}
A forward calculation of neutron multiplicity observables (e.g., $R_1,R_2,R_3$ and the derived Feynman variances $Y_2,Y_3$) requires specification of the neutron source term, including the SNM isotopic composition and any associated spontaneous-fission and $(\alpha,n)$ emission rates. In the present study, these quantities are fixed by the BeRP-ball definition used to generate all synthetic data in GADRAS (i.e., the isotopics and mass are treated as known, consistent with the problem setup in which the presence of Pu has already been established). In a deployment setting where isotopics are not known a priori, one would either (i) jointly estimate isotopics and shielding composition (multi-task or hierarchical inference), or (ii) treat the isotopics as nuisance parameters and marginalize over plausible ranges when evaluating candidate shielding configurations.  Finally, demonstration of the ability to utilize machine learning techniques to infer the isotopics has been previously demonstrated both computationally as well as experimentally~\cite{khatiwada23}.

\paragraph{Remark on noise modeling for $Y_2$ and $Y_3$}
The reported uncertainties on $Y_2$ and $Y_3$ are induced by counting-statistics fluctuations in the underlying multiplicity moments (equivalently in the measured singles/doubles/triples rates that determine $R_1,R_2,R_3$). Thus, variations in $Y_2$ and $Y_3$ implicitly correspond to variations in the estimated $R_q$ (and their covariances) rather than independent perturbations of unrelated quantities. In our machine-learning experiments, we adopt an effective measurement-error model by perturbing $Y_2$ and $Y_3$ directly with relative Gaussian errors calibrated to experimental characterizations. This approach is appropriate when the downstream classifier consumes $(Y_2,Y_3)$ as the neutron feature vector and the goal is to reflect the net uncertainty of those reported summary statistics. A more mechanistic alternative (not pursued here) would be to sample fluctuations at the level of the raw multiplicity counts/rates, propagate them through the estimator for $(R_1,R_2,R_3)$, and then recompute $(Y_2,Y_3)$, preserving any induced correlation between $Y_2$ and $Y_3$.  

Experimental uncertainties in neutron multiplicity observables such as $Y_2$ and $Y_3$ are not universal constants, but depend on several factors including count time, detector efficiency, coincidence gate structure, source strength and isotopic composition, and the degree of shielding and neutron multiplication. The relative uncertainties adopted in this work (approximately 2\% for $Y_2$ and 20\% for $Y_3$) are representative of values reported in experimental neutron multiplicity studies using MC-15–type detectors for plutonium metal objects under moderate shielding and count times on the order of several hundred seconds~\cite{LAPR-2021-037525,ensslin1991principles}. In practice, heavier shielding, lower multiplication, or shorter count times would generally increase these uncertainties, particularly for higher-order moments. In the present study, these values are used as an effective noise model to assess the robustness of the classification framework rather than as fixed detector performance limits.  Finally, to provide additional understanding of the importance of the uncertainty in these quantities a parametric examination of the impact of the magnitude of uncertainty is presented in Section 5.2.

% ===================== PATCH END =====================
Because the variance of multiplicity estimators scales inversely with the number of detected events, increasing the count time would be expected to reduce the uncertainty in $Y_2$ approximately as $1/\sqrt{T}$, while substantially longer count times are required to achieve comparable fractional reductions in the uncertainty of $Y_3$.

\section{Methods}
\label{sec:methods}

\subsection{Problem Statement}

We formulate our material identification problem as follows: Given neutron and spectroscopic measurements, ${\bf y}_{n}$ and ${\bf y}_{\rm spec}$ respectively, for a spherical object with an onion-like structure given by a sequence of radii of the nested shells $\{R_i\}_{i=1}^N$, identify the correct sequence of materials ${\bf \mu}=\{\mu_1,\dots\mu_N\}$ for each shell from a list of possible materials, $\mathcal{M}$,
e.g., $\mathcal{M} = \{{\rm Al}, {\rm Cu}, \dots, {\rm W} \}$.

\subsection{Multi-class classification framework}
\label{sec:multiclass}

The material identification task described in Section~\ref{sec:problem} can be cast as a supervised multi-class classification problem. For a given spherical configuration with $N$ concentric shells, let
\begin{equation}
    \boldsymbol{\mu} = \{\mu_1,\dots,\mu_N\} \in \mathcal{M}^N
\end{equation}
denote the ordered tuple of shell materials, where $\mathcal{M}$ is the discrete set of candidate materials (e.g., $\mathcal{M} = \{{\rm Al},{\rm B},{\rm Be},\dots,{\rm W}\}$ for the single-shell case described in Section~\ref{sec:problem}). Each distinct choice of $\boldsymbol{\mu}$ defines a \emph{class}. For fixed shell radii $\{R_i\}_{i=1}^N$, the total number of possible classes is
\begin{equation}
    K = |\mathcal{M}|^N \;,
\end{equation}
and we assume that the true configuration corresponds to one of these $K$ classes.

Rather than posing material identification as a residual-minimization problem---e.g., minimizing an $\ell_1$ or $\ell_2$ norm between measured and predicted observables---we adopt a supervised multi-class classification framework in which each admissible shielding-material configuration defines a discrete class. This choice is motivated by three considerations. First, the unknown shielding configuration is inherently discrete and combinatorial (particularly in the two-shell case), and a global $\ell_p$ metric imposes an artificial notion of geometric proximity on the hypothesis space that is not physically meaningful. Second, the feature vector combines heterogeneous observables (Poisson-dominated $\gamma$-ray photopeak counts and multiplicity-derived summaries such as $Y_2$ and $Y_3$) with markedly different dynamic ranges and uncertainty structures, making a single, globally valid residual norm nontrivial without a validated full covariance model. Third, practical deployments inevitably exhibit model mismatch (e.g., imperfect detector response, background, scattering, or source-term uncertainty), for which fixed $\ell_p$ objectives can be brittle. By training on noise-augmented synthetic data, the classifier implicitly learns decision boundaries that approximate the Bayes-optimal rule $\hat{\mu}=\arg\max_{\mu} p(\mu\mid \mathbf{x})$, thereby providing robustness to non-Gaussian noise, feature correlations, and forward-model uncertainty. In this regime, classification yields stable and interpretable material-identification performance, particularly for multi-layer configurations where ordering effects and feature degeneracies are pronounced.

\paragraph{Feature construction}
For each configuration we have two measurement modalities:
\begin{enumerate}
    \item A gamma-spectroscopic measurement $\mathbf{y}_{\rm spec} \in \mathbb{R}^{d_\gamma}$  of the HPGe spectrum (Section~\ref{sec:dataprep}). In this work we use the net counts in the 11 Pu-239 photo-peaks centered at
    \begin{equation}
    \begin{split}
        E_\ell \in \{&129.27, 203.53, 332.82, 344.99, 375.03, 392.91,\\
        &413.69, 451.46, 639.94, 645.87, 769.24\}\ {\rm keV} \;,
    \end{split}
    \end{equation}
    as determined in Refs.~\cite{korobkin2024isotopic,khatiwada2023machine}.
    \item A neutron multiplicity measurement $\mathbf{y}_{n}$ summarized by the Feynman variances $Y_2$ and $Y_3$ (Section~\ref{sec:dataprep}), i.e.
    \begin{equation}
        \mathbf{y}_{n} = (Y_2, Y_3)^\top \in \mathbb{R}^2 \;.
    \end{equation}
\end{enumerate}
These are combined into a single feature vector
\begin{equation}
\begin{split}
    \mathbf{x} &= \phi(\mathbf{y}_{\rm spec}, \mathbf{y}_{n}) \\
    &= \big( y_{\rm spec,1},\dots,y_{\rm spec,11},Y_2,Y_3 \big)^\top 
    \in \mathbb{R}^{d},\quad d = 13 \;,
\end{split}
\end{equation}
for the full multi-modal case. For modality-restricted experiments we use the appropriate sub-vector (gamma-only or gamma+$Y_2$).

To improve numerical conditioning and to balance the relative scale of gamma and neutron features, we normalize component-wise using the noiseless (baseline) simulated data. Let $\{\mathbf{x}_j^{(0)}\}_{j=1}^{K}$ denote the baseline feature vectors produced by GADRAS for each of the $K$ material configurations at fixed shell radii, without noise. For each feature index $m \in \{1,\dots,d\}$ we define
\begin{equation}
    s_m = \max_{1\le j\le K} x_{j,m}^{(0)} \;,
\end{equation}
and construct normalized features
\begin{equation}
    \tilde{\mathbf{x}} = (x_1/s_1,\dots,x_d/s_d)^\top \;,
\end{equation}
so that each feature component has maximum value equal to unity over the (baseline) training set.

\paragraph{Noisy observations and data augmentation}
Experiments were performed with a number of different acquisitions times using both the HPGe detector and the neutron multiplicity detector, as described in Section~\ref{sec:dataprep}. To mimic realistic measurement uncertainty we perturbed the baseline features using physically motivated noise models:
\begin{itemize}
    \item For each gamma feature (net counts in a given photo-peak) we draw a noisy realization from a Poisson distribution with mean equal to the baseline counts in that bin.
    \item For $Y_2$ and $Y_3$ we assume relative Gaussian errors consistent with experimental characterizations~\cite{LAPR-2021-037525}, namely
    \begin{align}
        \widehat{Y}_2 &= (1 + \mu_2)\, Y_2, \;\; \mu_2 \sim \mathcal{N}(0,\,\sigma_2^2), \;\; \sigma_2 = 2\times 10^{-2} \;, \\
        \widehat{Y}_3 &= (1 + \mu_3)\, Y_3, \;\; \mu_3 \sim \mathcal{N}(0,\,\sigma_3^2), \;\; \sigma_3 = 2\times 10^{-1}\;.
    \end{align}
\end{itemize}

To further examine the sensitivity of noise to the Feynman variances additional studies were performed in which the magnitude of the statistical variations were examined.

For each material configuration  one baseline feature vector was generated and then augmented t with  additional noisy realizations, as discussed below, obtained by independently sampling the above noise models. The test set consists of 10,000 independent noise realizations per configuration. Although the underlying baseline feature vectors are the same, the independent noise realizations provide a clear separation between training and testing distributions.  A motivation in varying the number of noise realizations in the training is to examine the impact of noise characterization i.e. when no additional noise samples are included in the training the noise characteristics are assumed to be poorly characterized whereas when many noise realizations are included the noise properties are well characterized.  This also serves as a first step in assessing the impact of model mismatch in the forward simulations and the test cases to be examined.

\paragraph{Multi-class random forest classifier}
Let $\mathcal{D} = \{(\tilde{\mathbf{x}}_i, y_i)\}_{i=1}^{N_{\rm train}}$ denote the training set at fixed shell radii $\{R_i\}_{i=1}^{N}$, where $y_i \in \{1,\dots,K\}$ encodes the material configuration class. We seek a classifier
\begin{equation}
    f: \mathbb{R}^d \rightarrow \{1,\dots,K\} \;,
\end{equation}
that approximates the Bayes optimal decision rule
\begin{equation}
    f^\star(\tilde{\mathbf{x}}) = \arg\max_{k \in \{1,\dots,K\}} \, p(y = k \mid \tilde{\mathbf{x}}) \;.
\end{equation}

We employ the Random Forest classifier, as implemented in the class \texttt{RandomForestClassifier} in scikit-learn~\cite{scikit-learn}. Closely related tree-based ensemble methods are discussed in~\cite{breiman1984classification}. A random forest is an ensemble of $T$ decision trees $\{h_t\}_{t=1}^{T}$ trained on bootstrap resamples of $\mathcal{D}$ with random feature sub-sampling. Each tree $h_t$ outputs a class probability vector
\begin{equation}
    \boldsymbol{\pi}^{(t)}(\tilde{\mathbf{x}}) 
    = \big(\pi^{(t)}_1(\tilde{\mathbf{x}}),\dots,\pi^{(t)}_K(\tilde{\mathbf{x}})\big) 
    \qquad \sum_{k=1}^K \pi^{(t)}_k(\tilde{\mathbf{x}}) = 1 \;.
\end{equation}
The forest-level posterior estimate is obtained by averaging over trees,
\begin{equation}
    \widehat{p}(y = k \mid \tilde{\mathbf{x}}) 
    = \frac{1}{T}\sum_{t=1}^{T} \pi^{(t)}_k(\tilde{\mathbf{x}})
    \qquad k = 1,\dots,K \;,
\end{equation}
and the predicted class is
\begin{equation}
    \hat{y} = f(\tilde{\mathbf{x}}) 
    = \arg\max_{k} \widehat{p}(y = k \mid \tilde{\mathbf{x}}) \;.
\end{equation}
This formulation naturally supports multi-class problems, handles heterogeneous feature scales after normalization, and is robust to modest levels of noise and feature correlation, which is essential for the present application where gamma and neutron features are jointly informative.

\paragraph{Per-radius training and evaluation}
Because the shell radii $\{R_i\}$ can be independently estimated with high precision from radiography (Section~\ref{sec:problem}), we condition on a discrete grid of plausible radius configurations and train a separate classifier for each point on this grid. For a fixed radius configuration, the training and testing protocol is:
\begin{enumerate}
    \item Generate baseline feature vectors for all $K$ material combinations using GADRAS.
    \item Normalize features using the procedure above and construct $N_{\rm train}$ augmented training samples via noise injection.
    \item Train a random forest classifier on $\mathcal{D}$.
    \item Generate $N_{\rm test}$ noisy test samples (10,000 per class) and compute performance metrics such as overall accuracy
    \begin{equation}
        {\rm Acc} = \frac{1}{N_{\rm test}} \sum_{i=1}^{N_{\rm test}} \mathbf{1}\{ \hat{y}_i = y_i \} \;,
    \end{equation}
    and confusion matrices to characterize the structure of the remaining misclassifications.
\end{enumerate}
In Section~\ref{sec:results} we report these metrics for both single-shell and double-shell configurations, and we quantify the incremental benefit of incorporating $Y_2$ and $Y_3$ relative to using gamma spectroscopy alone.

%\subsection{Alternative classification approaches}
%\label{sec:alt_approaches}

% --- Bridge from Section 4.2 (multi-class framework) into Section 4.3 ---
%To this point, we have described the supervised multi-class formulation and the random-forest implementation used for the primary results reported in Section~\ref{sec:results}. Because the feature space is low-dimensional and tabular (photopeak counts and multiplicity summaries) and because performance can depend on how a learner handles nonlinear boundaries and noise, we also performed an initial cross-algorithm comparison to verify that the observed multi-modality gains are not specific to a single classifier.

\subsection{Comparative evaluation of alternative classification approaches}
\label{sec:alt_approaches}

Although the primary results in this paper are reported using random forests, we also conducted an initial comparative study to assess the sensitivity of material-identification performance to the choice of classification algorithm. The goal of this study was not exhaustive hyperparameter optimization, but rather to determine whether the observed benefits of multi-modal features are specific to a single learner or persist across a range of commonly used classifiers for tabular, low-dimensional feature vectors. The algorithms evaluated in this study are summarized in Table~\ref{tab:single_shell_summary_compare}.
\begin{table}[htbp]
\centering
\caption{Single-shell material identification performance of different ML approaches, including random forest (RF), XGBoost, K-nearest neighbors (KNNs), support vector machine (SVM), and multilayer perceptrons (MLPs). We consider a 60s time window, a 0.5 inch single shell configuration, Gamma+$Y_2$+$Y_3$ as feature and different number of noisy samples per training simulations. $N=10{,}000$ independent noise realizations per feature set and thickness are generated for testing. Boldface indicates the best (highest) test accuracy within each method.}
\resizebox{0.5\columnwidth}{!}{%
\begin{tabular}{|c|c|c|c|}
\hline
\textbf{Method} &
\textbf{\# Noisy samples} &
\textbf{Train Acc.} &
\textbf{Test Acc.} \\
\hline
\hline

\multirow{5}{*}{RF}
  & 0 (clean data)
  & 1.000
  & 0.995 \\ \cline{2-4}
  & 2
  & 1.000
  & 0.993 \\ \cline{2-4}
  & 10
  & 1.000
  & 0.999 \\ \cline{2-4}
  & 100
  & 1.000
  & \textbf{1.000} \\ \cline{2-4}
  & 1000
  & 1.000
  & \textbf{1.000} \\
\hline
\hline

\multirow{5}{*}{XGBoost}
& 0 (clean data)
  & 0.048
  & 0.048 \\ \cline{2-4}
  & 2
  & 1.000
  & 0.786 \\ \cline{2-4}
  & 10
  & 1.000
  & 0.991 \\ \cline{2-4}
  & 100
  & 1.000
  & 0.996 \\ \cline{2-4}
  & 1000
  & 1.000
  & \textbf{1.000} \\
\hline
\hline

\multirow{5}{*}{KNNs}
& 0 (clean data)
  & 1.000
  & 0.961 \\ \cline{2-4}
  & 2
  & 1.000
  & 0.946 \\ \cline{2-4}
  & 10
  & 1.000
  & 0.964 \\ \cline{2-4}
  & 100
  & 1.000
  & 0.978 \\ \cline{2-4}
  & 1000
  & 1.000
  & \textbf{0.991} \\
\hline
\hline

\multirow{5}{*}{SVM}
& 0 (clean data)
  & 1.000
  & 0.960 \\ \cline{2-4}
  & 2
  & 1.000
  & 0.965 \\ \cline{2-4}
  & 10
  & 1.000
  & 0.997 \\ \cline{2-4}
  & 100
  & 1.000
  & \textbf{1.000} \\ \cline{2-4}
  & 1000
  & 1.000
  & \textbf{1.000} \\
\hline
\hline

\multirow{5}{*}{MLPs}
& 0 (clean data)
  & 1.000
  & 0.801 \\ \cline{2-4}
  & 2
  & 1.000
  & 0.829 \\ \cline{2-4}
  & 10
  & 0.914
  & 0.870 \\ \cline{2-4}
  & 100
  & 0.960
  & 0.951 \\ \cline{2-4}
  & 1000
  & 1.000
  & \textbf{1.000} \\
\hline

\end{tabular}
\label{tab:single_shell_summary_compare}
}
\end{table}
Specifically, we compared the following supervised multi-class classifiers:
(i) Random Forests (RF), which serve as the primary baseline in this work due to their robustness to heterogeneous feature scaling, nonlinear decision boundaries, and moderate noise;
(ii) gradient-boosted decision trees implemented using XGBoost (denoted XGBoost in Table~\ref{tab:single_shell_summary_compare});
(iii) support vector machines (SVM) with nonlinear kernels;
(iv) $k$-nearest neighbors (KNNs); and
(v) multilayer perceptrons (MLPs) as a representative neural-network model.

All methods were trained and evaluated under the same data-generation and noise-injection framework described in Sections~\ref{sec:dataprep}--\ref{sec:multiclass}. In particular, for each material configuration, a baseline (noiseless) feature vector was generated using GADRAS and the training set was augmented by drawing a specified number of independent noisy realizations per simulation (Table~\ref{tab:single_shell_summary_compare}). This augmentation procedure provides a controlled way to examine the impact of training-set enrichment on robustness to measurement variability (i.e., a proxy for model mismatch between idealized forward simulations and noisy observations). Performance was assessed using an independent test set of $N=10{,}000$ noisy realizations, consistent with the protocol used elsewhere in the paper.

The results in Table~\ref{tab:single_shell_summary_compare} show that multiple methods can achieve very high accuracy when sufficient noise augmentation is included in training, but the random forest classifier provides consistently strong performance with minimal tuning, and it reaches near-ceiling accuracy with comparatively modest augmentation. Based on these observations, random forests were selected as the primary classifier for the remainder of the study, and subsequent analyses focus on understanding the physics-driven structure of misclassifications and the incremental value of neutron multiplicity features rather than on algorithmic optimization.

%Table~4 presents a comparative evaluation of several classification algorithms for the single-shell 0.5~in configuration using gamma+$Y_2$+$Y_3$ features and a 60~s acquisition time. Methods examined include random forests, gradient-boosted trees (XGBoost), support vector machines, $k$-nearest neighbors, and multilayer perceptrons.

%The results in also demonstrate that all methods benefit from noise augmentation during training, and several achieve near-ceiling performance when sufficiently many noisy samples are included. However, random forests consistently achieve high accuracy with fewer noisy training samples and minimal tuning, motivating their use as the primary classifier in the remainder of this work.

\section{Results}
\label{sec:results}

In this section we evaluate the multi-class classification framework described in Section~\ref{sec:multiclass} for both single-shell and double-shell BeRP-ball configurations. For each fixed set of shell radii we train classifiers on normalized multi-modal feature vectors and assess performance on independent noisy realizations of the gamma-spectroscopic and neutron-multiplicity measurements. Results are reported in terms of classification accuracy, confusion matrices, and material-pair–resolved misclassification fractions, with particular emphasis on the impact of neutron multiplicity features, count time, and noise augmentation.

\subsection{Single-Shell Classification}

We first consider the case in which the BeRP ball is surrounded by a single shielding shell of unknown composition, with shell thickness assumed known from radiography. Table ~\ref{tab:single_shell_summary_thickness_material} summarizes classification performance for shell thicknesses ranging from 0.5 to 6.0~in using gamma-only features, gamma+$Y_2$, and gamma+$Y_2$+$Y_3$, assuming a 60~s acquisition time and  noise-free training data.
\begin{table}[htbp]
\centering
\small
\caption{Single-shell material identification performance versus shell thickness for three feature sets (Gamma only, Gamma+$Y_2$, Gamma+$Y_2$+$Y_3$) using 60s time window, clean data for training, and $N=10{,}000$ independent noise realizations per feature set and thickness for testing. Boldface indicates the best (highest) test accuracy within each thickness (ties bolded).}
\resizebox{0.55\columnwidth}{!}{%
\begin{tabular}{|c|c|c|c|}
\hline
\textbf{Thickness(in)} &
\textbf{Features} &
\textbf{Train Acc.} &
\textbf{Test Acc.} \\
\hline
\hline

\multirow{3}{*}{0.5}
  & Gamma
  & 1.000
  & 0.979 \\ \cline{2-4}
  & Gamma + $Y_2$
  & 1.000
  & \textbf{0.995} \\ \cline{2-4}
  & Gamma + $Y_2$ + $Y_3$
  & 1.000
  & \textbf{0.995} \\
\hline
\hline

\multirow{3}{*}{1.0}
  & Gamma
  & 1.000
  & 0.979 \\ \cline{2-4}
  & Gamma + $Y_2$
  & 1.000
  & 0.999 \\ \cline{2-4}
  & Gamma + $Y_2$ + $Y_3$
  & 1.000
  & \textbf{1.000} \\
\hline
\hline

\multirow{3}{*}{1.5}
  & Gamma
  & 1.000
  & 0.993 \\ \cline{2-4}
  & Gamma + $Y_2$
  & 1.000
  & 0.995 \\ \cline{2-4}
  & Gamma + $Y_2$ + $Y_3$
  & 1.000
  & \textbf{0.997} \\
\hline
\hline

\multirow{3}{*}{3.0}
  & Gamma
  & 1.000
  & 0.992 \\ \cline{2-4}
  & Gamma + $Y_2$
  & 1.000
  & 0.999 \\ \cline{2-4}
  & Gamma + $Y_2$ + $Y_3$
  & 1.000
  & \textbf{1.000}\\
\hline
\hline

\multirow{3}{*}{6.0}
  & Gamma
  & 1.000
  & 0.931 \\ \cline{2-4}
  & Gamma + $Y_2$
  & 1.000
  & 0.952 \\ \cline{2-4}
  & Gamma + $Y_2$ + $Y_3$
  & 1.000
  & \textbf{0.999} \\
\hline
\end{tabular}
\label{tab:single_shell_summary_thickness_material}
}
\end{table}

Across all thicknesses, high classification accuracy is achieved, even when only gamma-spectroscopic features are used. The most challenging configurations correspond to the thinnest (0.5~in) and thickest (6.0~in) shells. In the thin-shell limit, reduced interaction probability limits the contrast between materials, while in the thick-shell limit, strong attenuation suppresses usable spectral information. In both regimes, the inclusion of neutron multiplicity features improves performance, with gamma+$Y_2$ and gamma+$Y_2$+$Y_3$ consistently outperforming gamma-only classification.

The influence of acquisition time is examined in Table ~\ref{tab:single_shell_summary_compare_windows} for the 0.5~in shell configuration using gamma+$Y_2$+$Y_3$ features. As expected, increasing the count time improves classification accuracy by reducing statistical uncertainty in both the gamma and neutron measurements. Even at short acquisition times (15–60~s), classification performance remains high, demonstrating robustness of the approach under limited counting statistics.

\begin{table}[htbp]
\centering
\small
\caption{Single-shell material identification performance using random forest and diffenrent time windows. We consider the 0.5 inch shell configuration, Gamma+$Y_2$+$Y_3$ as feature and clean data for training. $N=10{,}000$ independent noise realizations per feature set and thickness are generated for testing. Boldface indicates the best (highest) test accuracy (ties bolded).}
\resizebox{0.35\columnwidth}{!}{%
\begin{tabular}{|c|c|c|}
\hline
\textbf{Time window} &
\textbf{Train Acc.} &
\textbf{Test Acc.} \\
\hline
\hline
  3600s
  & 1.000
  & \textbf{1.000} \\ \cline{1-3}
  60s
  & 1.000
  & 0.995 \\ \cline{1-3}
  15s
  & 1.000
  & 0.979 \\
\hline
\end{tabular}
\label{tab:single_shell_summary_compare_windows}
}
\end{table}

The impact of noise augmentation during training is shown in Table \ref{tab:single_shell_summary_compare_noisy}. When only noise-free data are used for training, test accuracy is slightly reduced, indicating sensitivity to measurement noise. As the number of noisy training samples per simulation increases, test performance improves and eventually saturates, with little additional gain beyond approximately 100 noisy realizations per baseline configuration. This behavior is consistent with variance reduction through noise-aware training.

\begin{table}[htbp]
\centering
\small
\caption{Single-shell material identification performance using random forest and 60s time window while considering the 0.5 inch shell configuration, Gamma+$Y_2$+$Y_3$ as feature and either clean data or different number of noisy samples per training simulations (note that when using noisy samples, the clean data are not included in the training). $N=10{,}000$ independent noise realizations per feature set and thickness are generated for testing. Boldface indicates the best (highest) test accuracy (ties bolded).}
\resizebox{0.35\columnwidth}{!}{%
\begin{tabular}{|c|c|c|}
\hline
\textbf{\# Noisy samples} &
\textbf{Train Acc.} &
\textbf{Test Acc.} \\
\hline
\hline
  0 (clean data)
  & 1.000
  & 0.995 \\ \cline{1-3}
  1
  & 1.000
  & 0.980 \\ \cline{1-3}
  2
  & 1.000
  & 0.993 \\ \cline{1-3}
  10
  & 1.000
  & 0.999 \\ \cline{1-3}
  100
  & 1.000
  & \textbf{1.000} \\ \cline{1-3}
  1000
  & 1.000
  & \textbf{1.000} \\
\hline
\end{tabular}
\label{tab:single_shell_summary_compare_noisy}
}
\end{table}

Material-resolved classification errors for the single-shell case are summarized in Figure \ref{fig:single_shell_summary}. Errors are concentrated in a small subset of materials and thicknesses, with low-$Z$ materials dominating the residual misclassifications at small thickness.

\begin{figure}[htbp]
    \centering % Centers the figure
    \includegraphics[width=0.6\columnwidth]{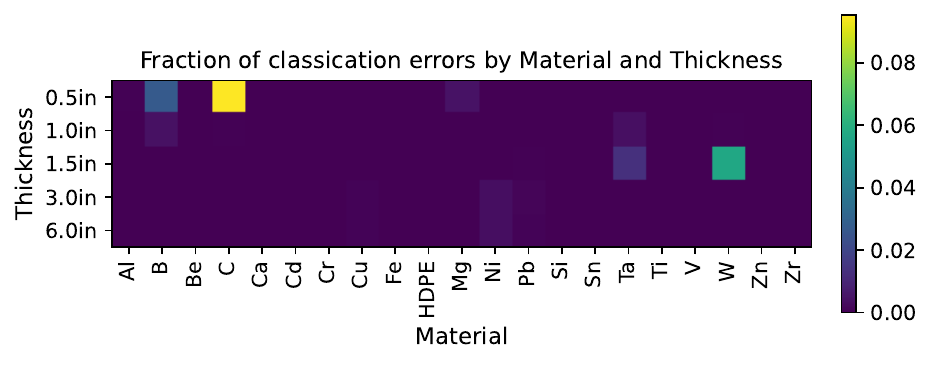} 
    \caption{Classification errors by material and thickness in single shell configuration using Gamma+$Y_2$+$Y_3$ as feature.} 
    \label{fig:single_shell_summary} % 
\end{figure}

Confusion matrices for representative single-shell cases are shown in Figure \ref{fig:single_shell_confusion_all} (note that these matrices are normalized so that the sum of classification errors for each reference material is 1 (if they are non-zeros)). For the 0.5~in shell, gamma-only classification exhibits systematic confusion between materials with similar attenuation properties, most notably carbon and boron. The addition of the $Y_2$ multiplicity feature substantially reduces these confusions by introducing sensitivity to neutron moderation and absorption. Including $Y_3$ produces only marginal additional changes, consistent with its comparatively larger uncertainty. For the 3.0~in shell, near-perfect classification is achieved even with gamma-only features, and neutron multiplicity provides little incremental benefit.

\begin{figure*}[t]
    \centering
    
    %-------------------- Row 1: shell thickness 0.5 in --------------------%
    \resizebox{1\columnwidth}{!}{%
    \begin{subfigure}[t]{0.35\linewidth}
        \centering
        \includegraphics[width=\linewidth]{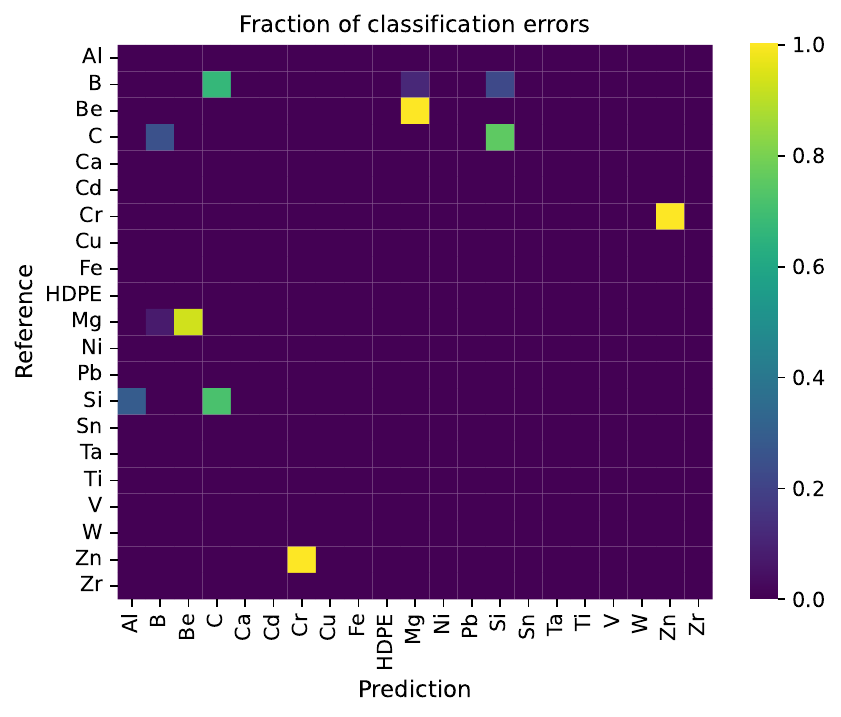}
        \caption{0.5in, Gamma}
        \label{fig:conf_gamma_0p5}
    \end{subfigure}
    \hfill
    \begin{subfigure}[t]{0.35\linewidth}
        \centering
        \includegraphics[width=\linewidth]{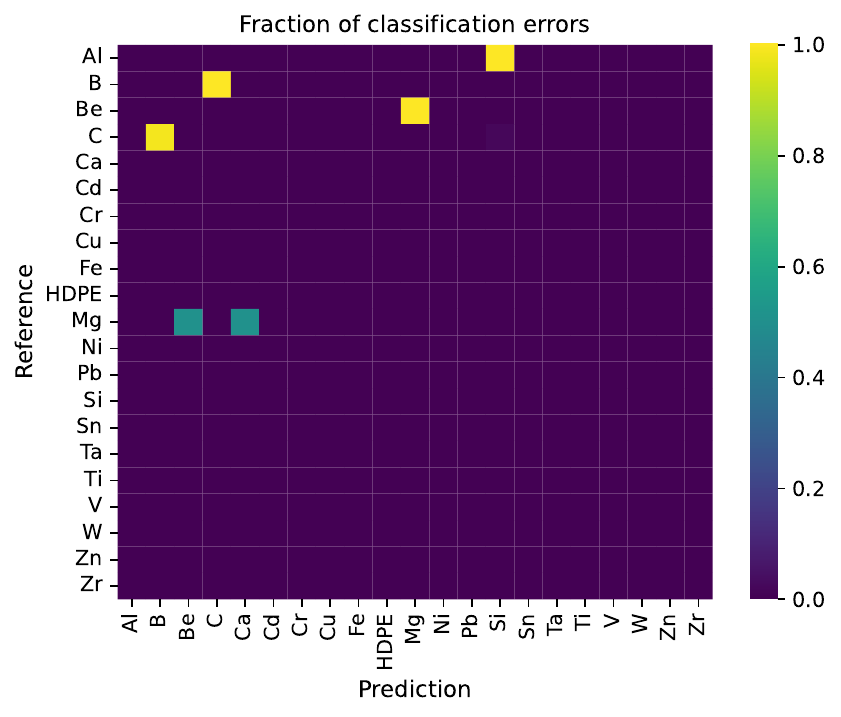}
        \caption{0.5in, Gamma+$Y_2$}
        \label{fig:conf_gamma_y2_0p5}
    \end{subfigure}
    \hfill
    \begin{subfigure}[t]{0.35\linewidth}
        \centering
        \includegraphics[width=\linewidth]{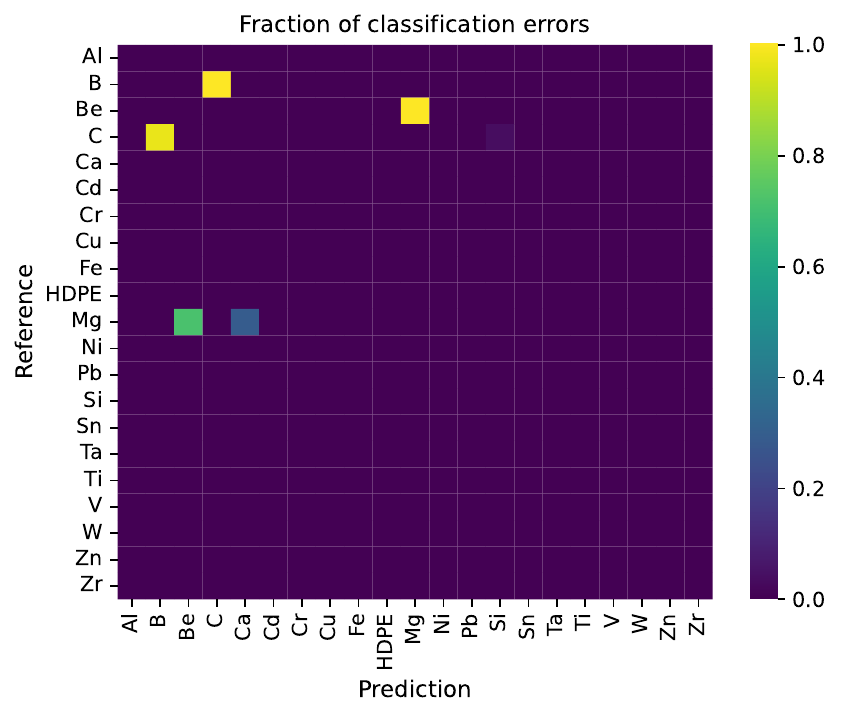}
        \caption{0.5in, Gamma + $Y_2$ + $Y_3$}
        \label{fig:conf_gamma_y2_y3_0p5}
    \end{subfigure}
    }
    \vspace{0.1cm}
    
    %-------------------- Row 2: shell thickness 3.0 in --------------------%
    \resizebox{1\columnwidth}{!}{%
    \begin{subfigure}[t]{0.35\linewidth}
        \centering
        \includegraphics[width=\linewidth]{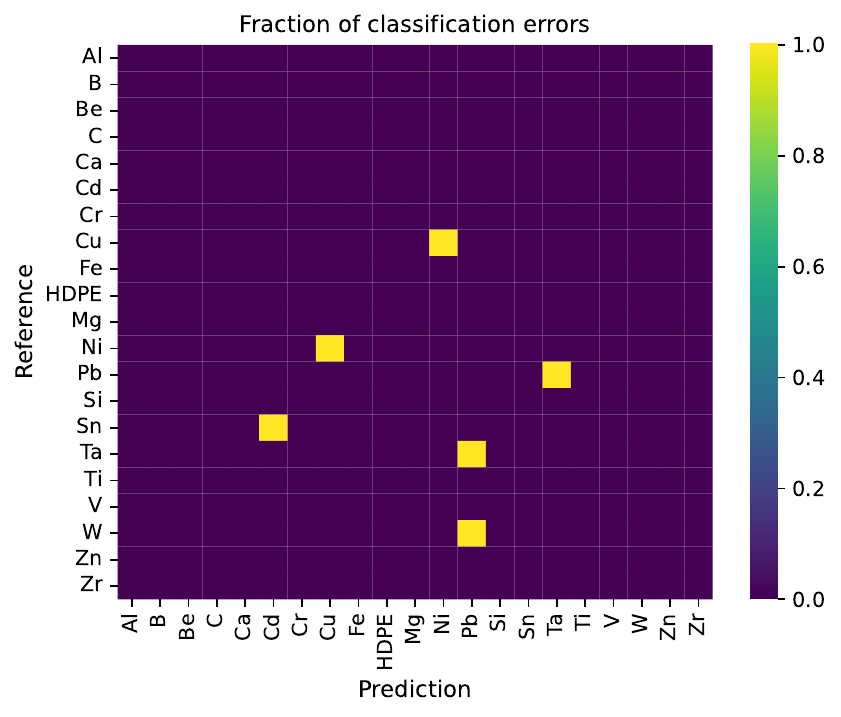}
        \caption{3.0in, Gamma}
        \label{fig:conf_gamma_3p0}
    \end{subfigure}
    \hfill
    \begin{subfigure}[t]{0.35\linewidth}
        \centering
        \includegraphics[width=\linewidth]{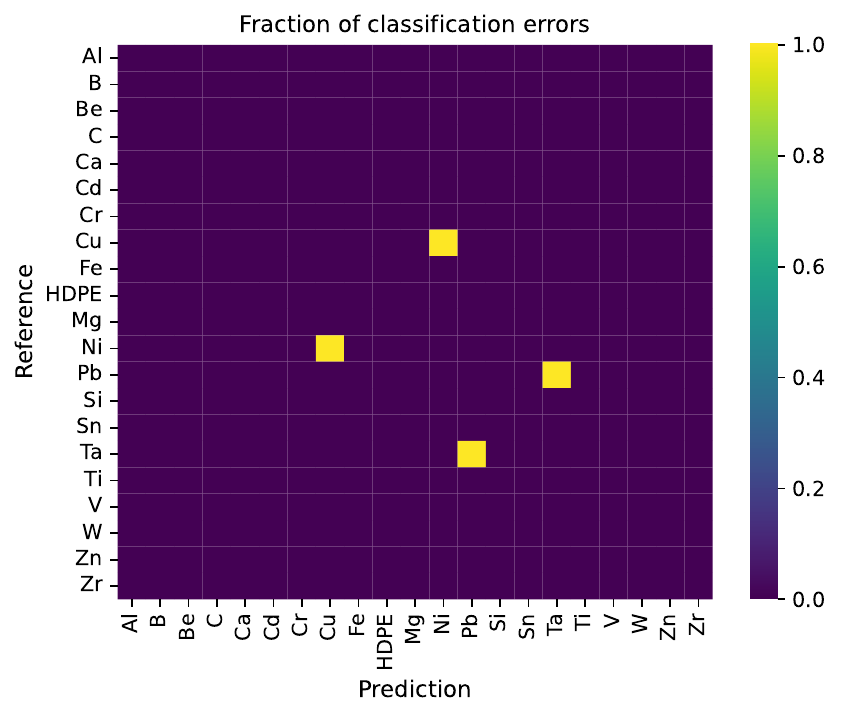}
        \caption{3.0in, Gamma + $Y_2$}
        \label{fig:conf_gamma_y2_3p0}
    \end{subfigure}
    \hfill
    \begin{subfigure}[t]{0.35\linewidth}
        \centering
        \includegraphics[width=\linewidth]{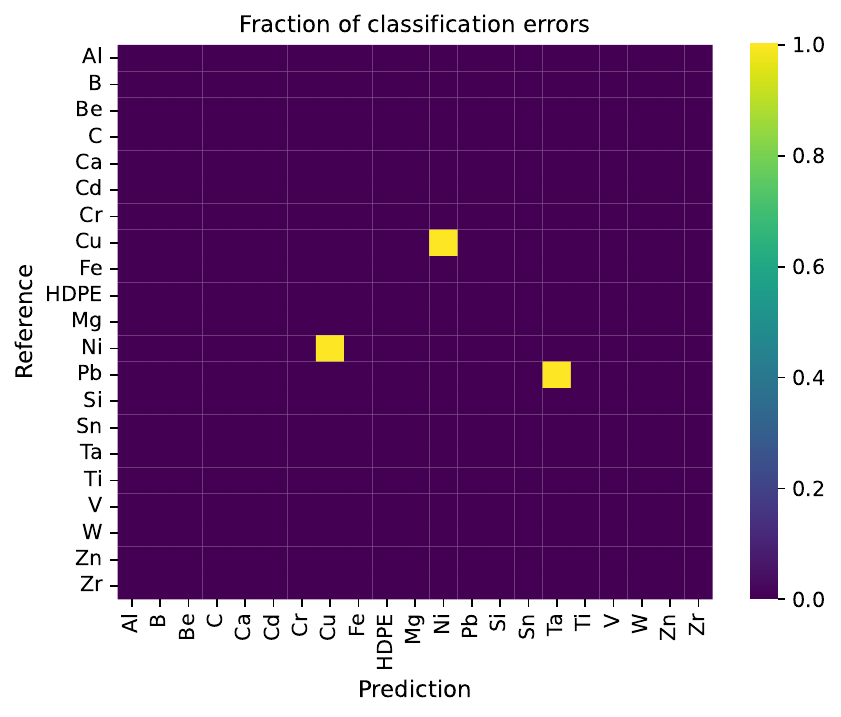}
        \caption{3.0in, Gamma + $Y_2$ + $Y_3$}
        \label{fig:conf_gamma_y2_y3_3p0}
    \end{subfigure}
    }
    \caption{Normalized confusion matrices for single-shell material identification at two shell thicknesses (rows) and three feature sets (columns): Gamma only (left), Gamma + $Y_2$ (middle), and Gamma + $Y_2$ + $Y_3$ (right), using 60s time window, clean data for training, and $N=10{,}000$ independent noise realizations per feature set and thickness for testing. Each panel shows the classification performance across all candidate shell materials under noisy measurement conditions. We normalize these matrices so that the sum of classification errors for each reference material is 1 (if they are non-zeros).}
    \label{fig:single_shell_confusion_all}
\end{figure*}

\subsection{Two-Shell Configurations}

We next examine the more challenging case in which the BeRP ball is surrounded by two concentric shells of unknown composition, with the two shell thicknesses (inner/outer) assumed known from radiography. Relative to the single-shell problem, the two-shell setting is substantially more difficult because the number of admissible classes grows combinatorially and, importantly, the measured spectrum depends on the \emph{ordering} of materials (i.e., which material is placed in the inner versus the outer shell).

%Table~\ref{tab:double_shell_summary_grid_ci} summarizes classification performance for representative thickness combinations using an 1800\,s acquisition time and clean (noise-free) training data. 

Table \ref{tab:double_shell_summary_grid_ci} summarizes classification performance for representative symmetric and asymmetric shell-thickness combinations using both 60s and 1800~s acquisition times with clean training data.  Several trends are evident. First, the double shell configurations with 60 sec count times suffers significant degradation in classification performance relative to the single shell cases, particularly for thin or highly asymmetric configurations.  This is anticipated both due to the increased number of admissible classes.  Second, adding the second-order Feynman variance $Y_2$ yields a large improvement in all cases for the 60s acquistion data.  Third, the inclusion of $Y_3$ provides a smaller incremental gain beyond gamma+$Y_2$.  Fourth, the increase in the count time to 1800 sec significantly improves the test accuracy.  Overall, these results indicate that neutron multiplicity measurements provide information that is complementary to gamma attenuation and is particularly valuable once multiple shielding layers are present.

% columnwidth
\begin{table*}[t]
\centering
% \small
\caption{Double-shell material identification performance for representative shell-thickness combinations (inner/outer) using either 60s or 1800s time window, and clean data for training. $N=10{,}000$ independent noise realizations per feature set and thickness are generated for testing. Boldface indicates the best (highest) test accuracy within each thickness combination (ties bolded).}
\resizebox{0.7\columnwidth}{!}{%
\begin{tabular}{|c|c|c|c|c|}
\hline
\textbf{Time window}&
\textbf{Thicknesses (in)} &
\textbf{Features} &
\textbf{Train Acc.} &
\textbf{Test Acc.} \\
\hline
\hline

\multirow{12}{*}{60s}
&
\multirow{3}{*}{0.5 / 0.5}
  & Gamma
  & 1.000
  & 0.671\\ \cline{3-5}
  & & Gamma + $Y_2$
  & 1.000
  & 0.815\\ \cline{3-5}
  & & Gamma + $Y_2$ + $Y_3$
  & 1.000
  & \textbf{0.870} \\
  \cline{2-5}%\\
  % \cline{2-5}
% \\ \cline{1-5}
% \\ \cline{1-5}

&
\multirow{3}{*}{3.0 / 3.0}
  & Gamma
  & 1.000
  & 0.687 \\ \cline{3-5}
  & & Gamma + $Y_2$
  & 1.000
  & 0.808 \\ \cline{3-5}
  & & Gamma + $Y_2$ + $Y_3$
  & 1.000
  & \textbf{0.840} \\
\cline{2-5}

&
\multirow{3}{*}{1.0 / 6.0}
  & Gamma
  & 1.000
  & 0.621 \\ \cline{3-5}
  & & Gamma + $Y_2$
  & 1.000
  & 0.717 \\ \cline{3-5}
  & & Gamma + $Y_2$ + $Y_3$
  & 1.000
  & \textbf{0.758} \\
\cline{2-5}

&
\multirow{3}{*}{6.0 / 1.0}
 & Gamma
  & 1.000
  & 0.648 \\ \cline{3-5}
  & & Gamma + $Y_2$
  & 1.000
  & 0.745 \\ \cline{3-5}
  & & Gamma + $Y_2$ + $Y_3$
  & 1.000
  & \textbf{0.796} \\
\hline
\hline
\multirow{12}{*}{1800s}
&
\multirow{3}{*}{0.5 / 0.5}
  & Gamma
  & 1.000
  & 0.978 \\ \cline{3-5}
  & & Gamma + $Y_2$
  & 1.000
  & 0.989 \\ \cline{3-5}
  & & Gamma + $Y_2$ + $Y_3$
  & 1.000
  & \textbf{0.990} \\
  \cline{2-5}%\\
  % \cline{2-5}
% \\ \cline{1-5}
% \\ \cline{1-5}

&
\multirow{3}{*}{3.0 / 3.0}
  & Gamma
  & 1.000
  & 0.968 \\ \cline{3-5}
  & & Gamma + $Y_2$
  & 1.000
  & \textbf{0.981} \\ \cline{3-5}
  & & Gamma + $Y_2$ + $Y_3$
  & 1.000
  & \textbf{0.981} \\
\cline{2-5}

&
\multirow{3}{*}{1.0 / 6.0}
  & Gamma
  & 1.000
  & 0.923 \\ \cline{3-5}
  & & Gamma + $Y_2$
  & 1.000
  & 0.950 \\ \cline{3-5}
  & & Gamma + $Y_2$ + $Y_3$
  & 1.000
  & \textbf{0.964} \\
\cline{2-5}

&
\multirow{3}{*}{6.0 / 1.0}
 & Gamma
  & 1.000
  & 0.966 \\ \cline{3-5}
  & & Gamma + $Y_2$
  & 1.000
  & 0.982 \\ \cline{3-5}
  & & Gamma + $Y_2$ + $Y_3$
  & 1.000
  & \textbf{0.988} \\
\hline

\end{tabular}
\label{tab:double_shell_summary_grid_ci}
}
\end{table*}

A clear ordering dependence is observed when comparing the 1.0/6.0 and 6.0/1.0 configurations. With gamma-only features using 1800 sec count times, the 1.0/6.0 case attains 0.923 test accuracy while the reversed ordering achieves 0.966 (Table~\ref{tab:double_shell_summary_grid_ci}). This behavior is consistent with the directional nature of gamma transport through layered media: when the thicker shell is closer to the source, the spectrum can be strongly attenuated and spectrally shaped before traversing the outer shell, reducing the dynamic range available to disentangle the two materials at the detector. Incorporating neutron multiplicity mitigates (but does not eliminate) this asymmetry. In particular, gamma+$Y_2$+$Y_3$ improves the 1.0/6.0 case to 0.964 and the 6.0/1.0 case to 0.988, reflecting the fact that the neutron features encode integrated moderation/absorption properties of the full assembly and are therefore less sensitive to ordering than gamma attenuation alone.

The impact of counting time is quantified in Table~\ref{tab:two_shell_summary_compare_windows} for the 0.5/0.5 configuration using the full feature set (gamma+$Y_2$+$Y_3$) and clean training data. Performance improves monotonically with acquisition time: the test accuracy increases from 0.808 (15\,s) to 0.870 (60\,s), 0.968 (600\,s), and approaches a plateau near 0.99 for 1800--3600\,s. This trend is expected because longer acquisitions reduce counting-statistics uncertainty in both the gamma photopeak counts and the multiplicity-derived statistics, thereby shrinking class overlap in feature space.  It should be noted that for computational studies the inclusion of noise in the training data serves as a very effective means to mitigate the impact of noise in the test data.  The use of noise in these investigations may also be thought of as a proxy for model mismatch between the forward operators photon and neutron and the experimental data. It should be noted that this proxy does not address systematic mismatches that cannot be addressed via sampling.

\begin{table}[htbp]
\centering
\small
\caption{Double-shell material identification performance using random forest and diffenrent time windows. We consider the 0.5/0.5 inch shell configuration, Gamma+$Y_2$+$Y_3$ as feature and clean data for training. $N=10{,}000$ independent noise realizations per feature set and thickness are generated for testing. Boldface indicates the best (highest) test accuracy (ties bolded).}
\resizebox{0.35\columnwidth}{!}{%
\begin{tabular}{|c|c|c|}
\hline
\textbf{Time window} &
\textbf{Train Acc.} &
\textbf{Test Acc.} \\
\hline
\hline
  3600s
  & 1.000
  & \textbf{0.992} \\ \cline{1-3}
  1800s
  & 1.000
  & 0.990 \\ \cline{1-3}
  600s
  & 1.000
  & 0.968  \\ \cline{1-3}
  60s
  & 1.000
  & 0.870 \\ \cline{1-3}
  15s
  & 1.000
  & 0.808 \\
\hline
\end{tabular}
\label{tab:two_shell_summary_compare_windows}
}
\end{table}

\begin{table}[htbp]
\centering
\small
\caption{Double-shell material identification performance using random forest with 60s or 1800s time window while considering the 0.5/0.5 inch shell configuration, Gamma+$Y_2$+$Y_3$ as feature and either clean data or different number of noisy samples per training simulations (note that when using noisy samples, the clean data are not included in the training). $N=10{,}000$ independent noise realizations per feature set and thickness  are generated for testing. Boldface indicates the best (highest) test accuracy (ties bolded).}
\resizebox{0.5\columnwidth}{!}{%
\begin{tabular}{|c|c|c|c|}
\hline
 \textbf{Time window}&
\textbf{\# Noisy samples} &
\textbf{Train Acc.} &
\textbf{Test Acc.} \\
\hline
\hline
\multirow{6}{*}{60s}
 & 0 (clean data)
  & 1.000
  & 0.870 \\ \cline{2-4}
 & 1 
  & 1.000
  & 0.731 \\ \cline{2-4}
 & 2 
  & 1.000
  & 0.875 \\ \cline{2-4}
 & 10
  & 1.000
  & 0.983 \\ \cline{2-4}
 & 100
  & 1.000
  & 0.993 \\ \cline{2-4}
 & 1000
  & 1.000
  & \textbf{0.995} \\ \cline{2-4}
& 10000
  & 1.000
  & \textbf{0.996} \\ \cline{2-4}
  & 100000
  & 1.000
  & \textbf{1.000}\\
\hline
\hline
\multirow{6}{*}{1800s}
 & 0 (clean data)
  & 1.000
  & 0.990 \\ \cline{2-4}
 & 1 
  & 1.000
  & 0.978 \\ \cline{2-4}
 & 2 
  & 1.000
  & 0.996 \\ \cline{2-4}
 & 10
  & 1.000
  & \textbf{1.000} \\ \cline{2-4}
 & 100
  & 1.000
  & \textbf{1.000} \\ \cline{2-4}
 & 1000
  & 1.000
  & \textbf{1.000} \\
\hline
\end{tabular}
\label{tab:two_shell_summary_compare_noisy}
}
\end{table}
Table~\ref{tab:two_shell_summary_compare_noisy} shows the impact of augmenting the training set with noisy realizations (here for the 0.5/0.5 case at 1800\,s using gamma+$Y_2$+$Y_3$). Training on noise-free data alone achieves 0.990 test accuracy, while adding a small number of noisy realizations can initially reduce performance (e.g., 0.978 for 1 noisy sample/simulation), reflecting the fact that insufficient sampling of the noise distribution can introduce a mild mismatch between the learned decision boundaries and the test-time noise variability. As the number of noisy realizations increases, performance improves and saturates: with 10 or more noisy samples per simulation, test accuracy reaches 1.000 and remains at that level for 100 and 1000 samples. This behavior is consistent with the interpretation that noise augmentation primarily improves robustness by better sampling the effective measurement-error distribution.

While the aggregate classification accuracies in Table \ref{tab:double_shell_summary_grid_ci} quantify the overall impact of shell ordering, they do not reveal which specific material combinations dominate the residual errors. To better understand the physical origin of the remaining misclassifications and the role of neutron multiplicity in mitigating them, we  examine the material-pair–resolved error distributions for the two-shell configurations, as summarized in Figures \ref{fig:errors_2shell_symmetric} and \ref{fig:errors_2shell_asymmetric}.  In the gamma-only case, classification errors are broadly distributed across many material combinations, indicating substantial ambiguity in disentangling the contributions of multiple shielding layers. When neutron multiplicity information is incorporated through the inclusion of the $Y_2$ Feynman variance, the structure of the error matrices changes markedly: the majority of material pairs exhibit near-zero misclassification rates, with residual errors concentrated into a small subset of pairs. These remaining ambiguities predominantly involve materials with similar neutron moderation and absorption properties, rather than materials that are primarily degenerate under gamma attenuation alone. This demonstrates that neutron multiplicity measurements do not simply reduce noise in the classifier, but instead alter the geometry of the feature space by resolving gamma-degenerate directions.

 The qualitative behavior observed in Figures~6 and~7 can be formalized using a simple measure of error concentration. Let $E_{ij}$ denote the fraction of misclassified test cases for material pair $(i,j)$, normalized such that $\sum_{i,j} E_{ij} = 1$ over all error entries. A useful diagnostic is the entropy of this distribution,
\[
H = -\sum_{i,j} E_{ij} \log E_{ij} \;,
\]
which quantifies how broadly classification errors are distributed across material pairs. 
In this work, the entropy of the pairwise misclassification distribution is computed using natural logarithms. Accordingly, entropy values are reported in \emph{nats}, where one nat corresponds to an entropy computed with base-$e$ logarithms. 
Using natural logarithms is convenient in this context because it yields a direct interpretation of the effective number of error modes as $N_{\mathrm{eff}}=\exp(H)$.

In the gamma-only case, the error distribution exhibits high entropy, reflecting diffuse confusion among many material combinations. In contrast, the inclusion of neutron multiplicity—particularly the $Y_2$ feature—produces a markedly lower-entropy error distribution, in which misclassifications are concentrated into a small number of physically similar material pairs. This reduction in error entropy provides a compact quantitative interpretation of the visual collapse of errors observed in the misclassification matrices and underscores the role of neutron multiplicity in resolving gamma-degenerate ambiguities.

\begin{table*}[t]
\centering
\small
\caption{Double-shell material identification performance  for 0.5/0.5 inch shell configuration, with different level of noise using (Gamma only, Gamma+$Y_2$, Gamma+$Y_2$+$Y_3$) using either 60s or 1800s time window, clean data for training, and $N=10{,}000$ independent noise realizations per feature set and thickness for testing. Boldface indicates the best (highest) test accuracy within each thickness (ties bolded).}
\resizebox{0.7\columnwidth}{!}{%
\begin{tabular}{|c|c|c|c|c|}
\hline
\textbf{Time window} &
\textbf{Noise level} &
\textbf{Features} &
\textbf{Train Acc.} &
\textbf{Test Acc.} \\
\hline
\hline

\multirow{9}{*}{60s}
&
\multirow{3}{*}{$\sigma_2=0.02, \sigma_3=0.2$}
  & Gamma
  & 1.000
  & 0.671 \\ \cline{3-5}
  &
  & Gamma + $Y_2$
  & 1.000
  & 0.815 \\ \cline{3-5}
  &
  & Gamma + $Y_2$ + $Y_3$
  & 1.000
  & \textbf{0.870} \\
\cline{2-5}

&
\multirow{3}{*}{$\sigma_2=0.04, \sigma_3=0.4$}
  & Gamma
  & 1.000
  & 0.644 \\ \cline{3-5}
  &
  & Gamma + $Y_2$
  & 1.000
  &  0.805 \\ \cline{3-5}
  &
  & Gamma + $Y_2$ + $Y_3$
  & 1.000
  &  \textbf{0.847} \\
\cline{2-5}

&
\multirow{3}{*}{$\sigma_2=0.08, \sigma_3=0.8$}
  & Gamma
  & 1.000
  & 0.644  \\ \cline{3-5}
  &
  & Gamma + $Y_2$
  & 1.000
  & 0.776 \\ \cline{3-5}
  &
  & Gamma + $Y_2$ + $Y_3$
  & 1.000
  & \textbf{0.782} \\
\hline
\hline

\multirow{9}{*}{1800s}
&
\multirow{3}{*}{$\sigma_2=0.02, \sigma_3=0.2$}
  & Gamma
  & 1.000
  & 0.978 \\ \cline{3-5}
  &
  & Gamma + $Y_2$
  & 1.000
  & 0.989 \\ \cline{3-5}
  &
  & Gamma + $Y_2$ + $Y_3$
  & 1.000
  & \textbf{0.990} \\
\cline{2-5}

&
\multirow{3}{*}{$\sigma_2=0.04, \sigma_3=0.4$}
  & Gamma
  & 1.000
  & 0.974 \\ \cline{3-5}
  &
  & Gamma + $Y_2$
  & 1.000
  &  0.988 \\ \cline{3-5}
  &
  & Gamma + $Y_2$ + $Y_3$
  & 1.000
  &  \textbf{0.990} \\
\cline{2-5}

&
\multirow{3}{*}{$\sigma_2=0.08, \sigma_3=0.8$}
  & Gamma
  & 1.000
  & 0.967  \\ \cline{3-5}
  &
  & Gamma + $Y_2$
  & 1.000
  & 0.974 \\ \cline{3-5}
  &
  & Gamma + $Y_2$ + $Y_3$
  & 1.000
  & \textbf{0.984} \\
\hline
\end{tabular}
\label{tab:double_shell_noise}
}
\end{table*}

The impact of the noise levels on the classification performance is presented in Table \ref{tab:double_shell_noise}.  The selected count time of 1800 sec for data acquisition is sufficient to largely negate the impact of significant noise.  It should be noted however, if the origins of the model mismatch are systematic  potentially larger impacts on classification performance will be observed. To resolve this, validation through experiments is required.  Overall, these results demonstrate that neutron multiplicity measurements provide complementary information to gamma spectroscopy that is essential for robust material identification in multi-layer shielding scenarios.

%The asymmetric two-shell configurations further illustrate that ordering effects manifest at the level of individual material pairs. In particular, certain material combinations exhibit elevated misclassification probabilities only when a specific material is located in the inner shell, confirming that the observed asymmetry arises from the directional nature of radiation transport through layered media rather than from classifier artifacts. Finally, the addition of the third-order Feynman variance $Y_3$ produces only marginal changes in the error structure relative to gamma+$Y_2$. Given the comparatively large uncertainty assumed for $Y_3$, this behavior indicates that higher-order multiplicity moments provide limited additional discriminatory power in the present measurement regime and primarily redistribute residual errors rather than systematically reducing them.

\begin{table}[!htbp]
\centering
\small
\caption{Double-shell material identification performance with entropy using (Gamma only, Gamma+$Y_2$, Gamma+$Y_2$+$Y_3$) using 1800s time window, clean data for training, and $N=10{,}000$ independent noise realizations per feature set and thickness for testing. Boldface indicates the best (highest) test accuracy within each thickness (ties bolded).}
\resizebox{0.45\columnwidth}{!}{%
\begin{tabular}{|c|c|c|}
\hline
\textbf{Thickness (in)} &
\textbf{Features} &
\textbf{Entropy}  \\
\hline
\hline

\multirow{3}{*}{$0.5/0.5$}
  & Gamma 
  & 3.337 \\ \cline{2-3}
  & Gamma + $Y_2$
  & 2.622 \\ \cline{2-3}
  & Gamma + $Y_2$ + $Y_3$
  & \textbf{2.362} \\
\hline
\hline
\multirow{3}{*}{$3.0/3.0$}
  & Gamma 
  & 3.313 \\ \cline{2-3}
  & Gamma + $Y_2$
  & 2.672 \\ \cline{2-3}
  & Gamma + $Y_2$ + $Y_3$
  & \textbf{2.547} \\
\hline
\hline
\multirow{3}{*}{$1.0/6.0$}
  & Gamma 
  & 3.904 \\ \cline{2-3}
  & Gamma + $Y_2$
  & 3.556 \\ \cline{2-3}
  & Gamma + $Y_2$ + $Y_3$
  & \textbf{3.168} \\
\hline
\hline
\multirow{3}{*}{$6.0/1.0$}
  & Gamma 
  & 3.362 \\ \cline{2-3}
  & Gamma + $Y_2$
  & 3.092 \\ \cline{2-3}
  & Gamma + $Y_2$ + $Y_3$
  & \textbf{2.792} \\
\hline
\hline
\multirow{3}{*}{$0.5/6.0$}
  & Gamma 
  & 3.914 \\ \cline{2-3}
  & Gamma + $Y_2$
  & 3.629 \\ \cline{2-3}
  & Gamma + $Y_2$ + $Y_3$
  & \textbf{3.329} \\
\hline
\hline
\multirow{3}{*}{$6.0/3.0$}
  & Gamma 
  & 3.512 \\ \cline{2-3}
  & Gamma + $Y_2$
  & 3.082 \\ \cline{2-3}
  & Gamma + $Y_2$ + $Y_3$
  & \textbf{2.437} \\
\hline
\end{tabular}
\label{tab:double_shell_entropy}
}
\end{table}

\begin{figure*}[!htbp]
\centering

% ---------------- Row 1: 0.5 / 0.5 ----------------
\resizebox{1\columnwidth}{!}{%
\begin{subfigure}[t]{0.35\linewidth}
  \centering
  \includegraphics[width=\linewidth]{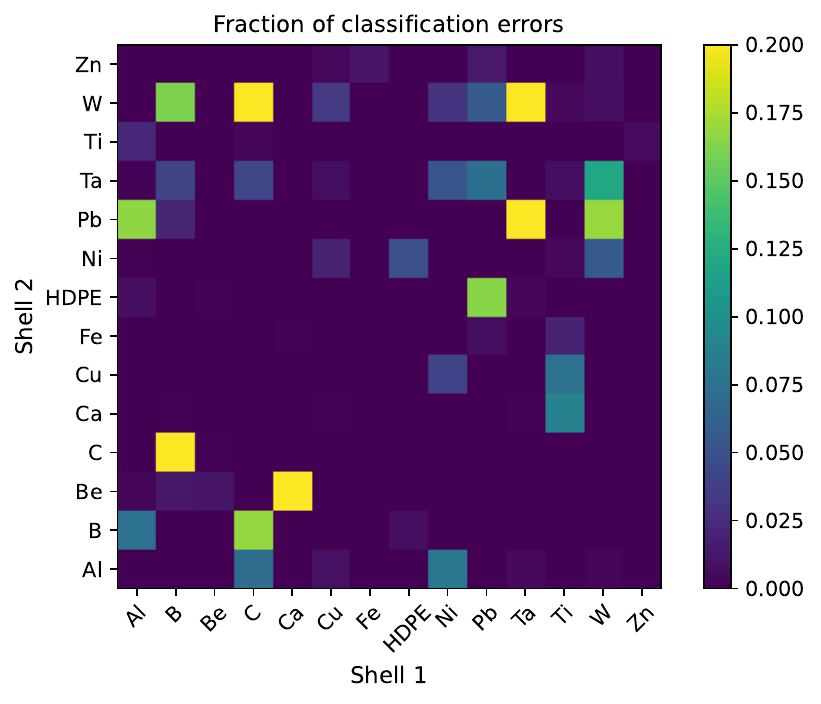}
  \caption{0.5/0.5, Gamma}
  \label{fig:err_0p5_0p5_gamma}
\end{subfigure}
\hfill
\begin{subfigure}[t]{0.35\linewidth}
  \centering
  \includegraphics[width=\linewidth]{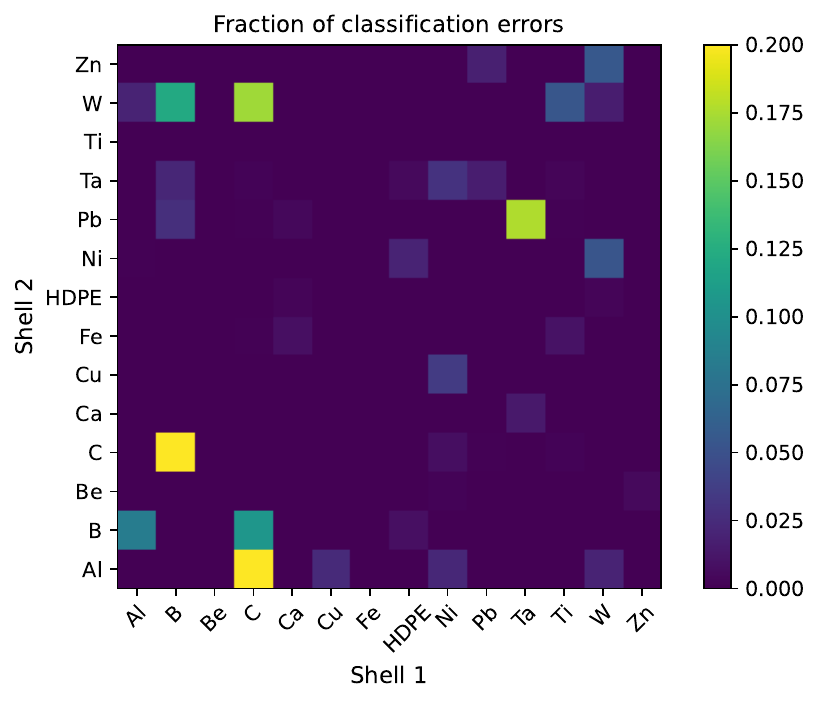}
  \caption{0.5/0.5, Gamma+$Y_2$}
  \label{fig:err_0p5_0p5_gammay2}
\end{subfigure}
\hfill
\begin{subfigure}[t]{0.35\linewidth}
  \centering
  \includegraphics[width=\linewidth]{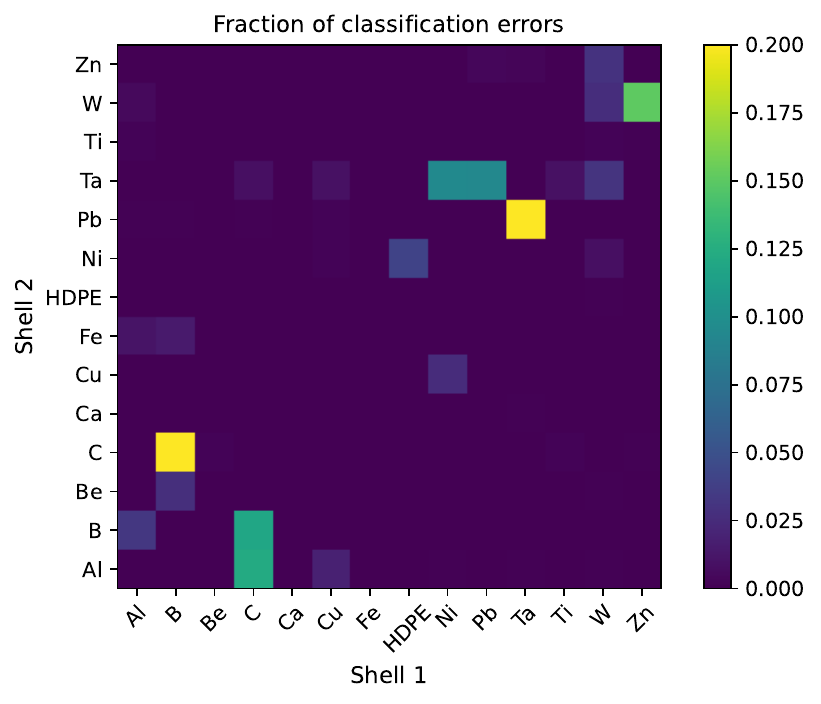}
  \caption{0.5/0.5, Gamma+$Y_2$+$Y_3$}
  \label{fig:err_0p5_0p5_gammay2y3}
\end{subfigure}
}
% \vspace{0.1cm}

% ---------------- Row 2: 3.0 / 3.0 ----------------
\resizebox{1\columnwidth}{!}{%
\begin{subfigure}[t]{0.35\linewidth}
  \centering
  \includegraphics[width=\linewidth]{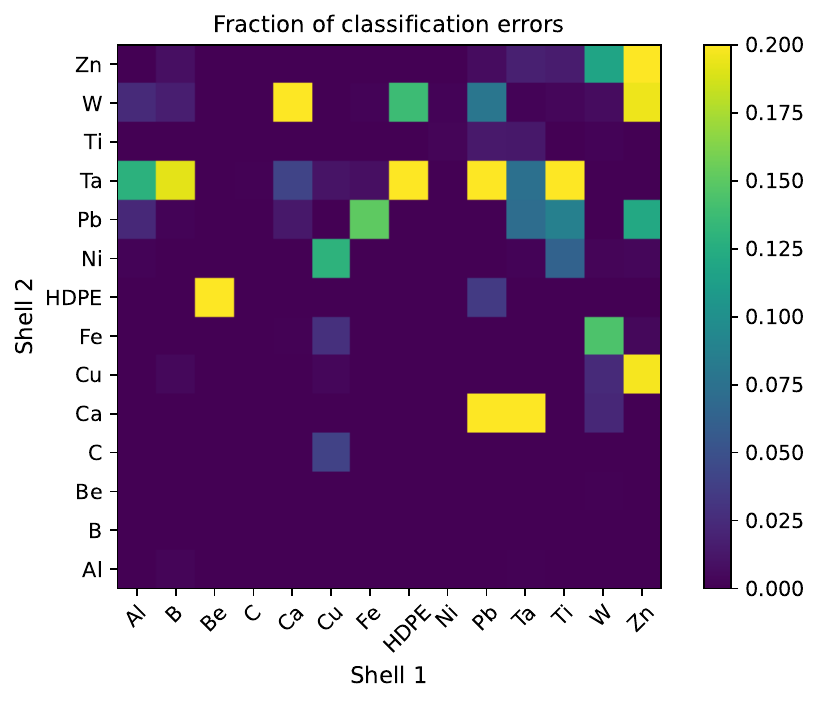}
  \caption{3.0/3.0, Gamma}
  \label{fig:err_3p0_3p0_gamma}
\end{subfigure}
\hfill
\begin{subfigure}[t]{0.35\linewidth}
  \centering
  \includegraphics[width=\linewidth]{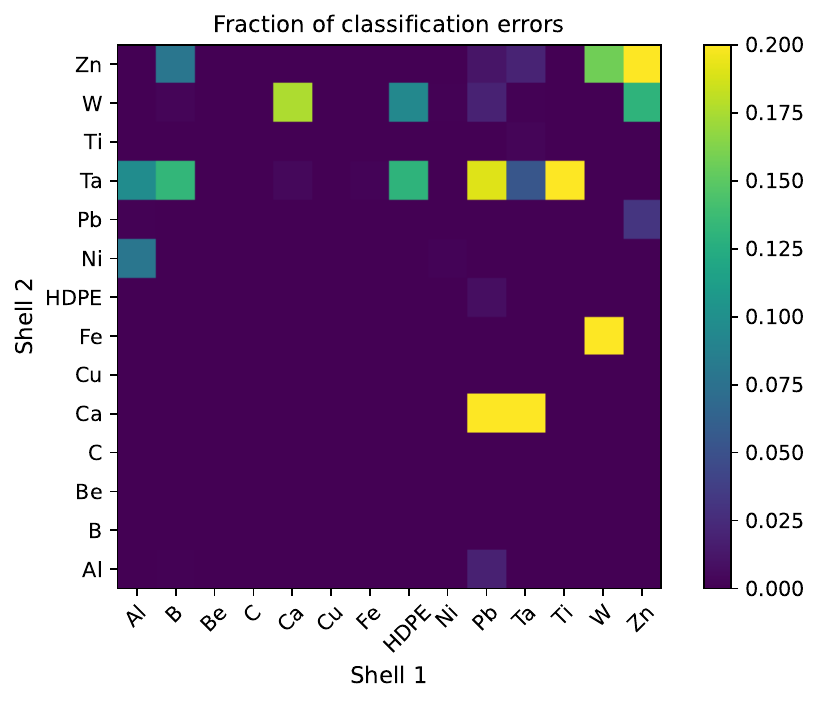}
  \caption{3.0/3.0, Gamma+$Y_2$}
  \label{fig:err_3p0_3p0_gammay2}
\end{subfigure}
\hfill
\begin{subfigure}[t]{0.35\linewidth}
  \centering
  \includegraphics[width=\linewidth]{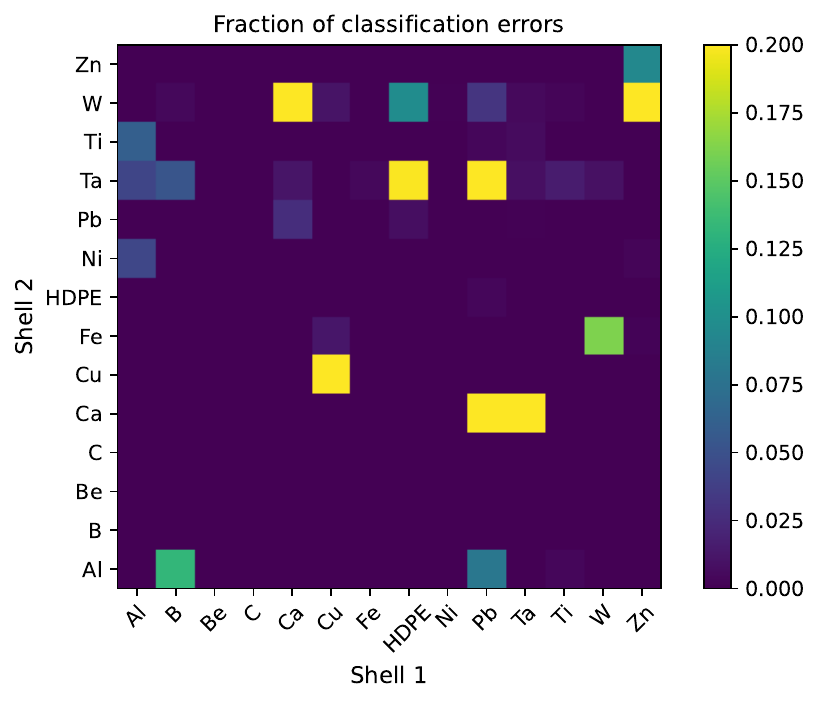}
  \caption{3.0/3.0, Gamma+$Y_2$+$Y_3$}
  \label{fig:err_3p0_3p0_gammay2y3}
\end{subfigure}
}
% \vspace{0.1cm}

% ---------------- Row 3: 1.0 / 6.0 ----------------
\resizebox{1\columnwidth}{!}{%
\begin{subfigure}[t]{0.35\linewidth}
  \centering
  \includegraphics[width=\linewidth]{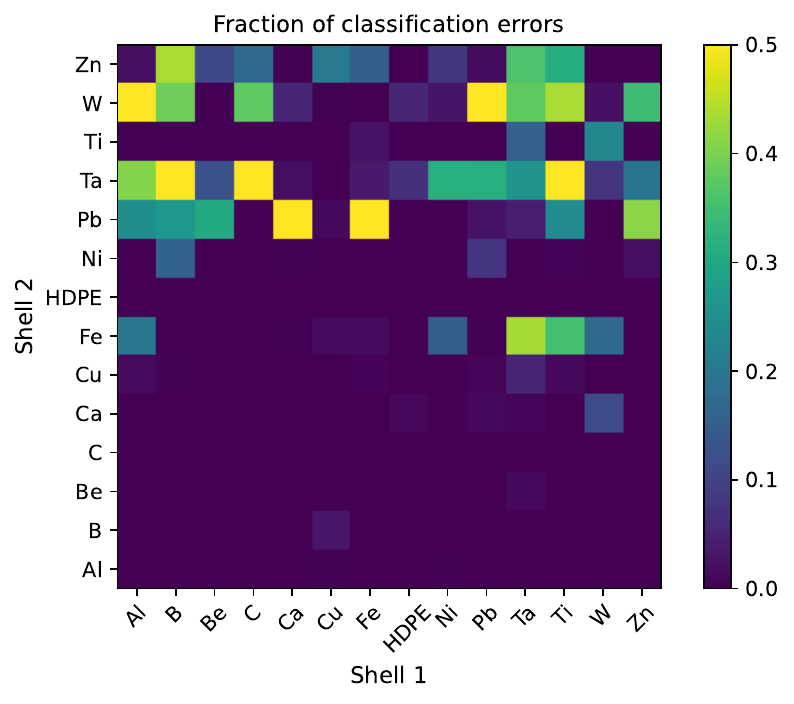}
  \caption{1.0/6.0, Gamma}
  \label{fig:err_1p0_6p0_gamma}
\end{subfigure}
\hfill
\begin{subfigure}[t]{0.35\linewidth}
  \centering
  \includegraphics[width=\linewidth]{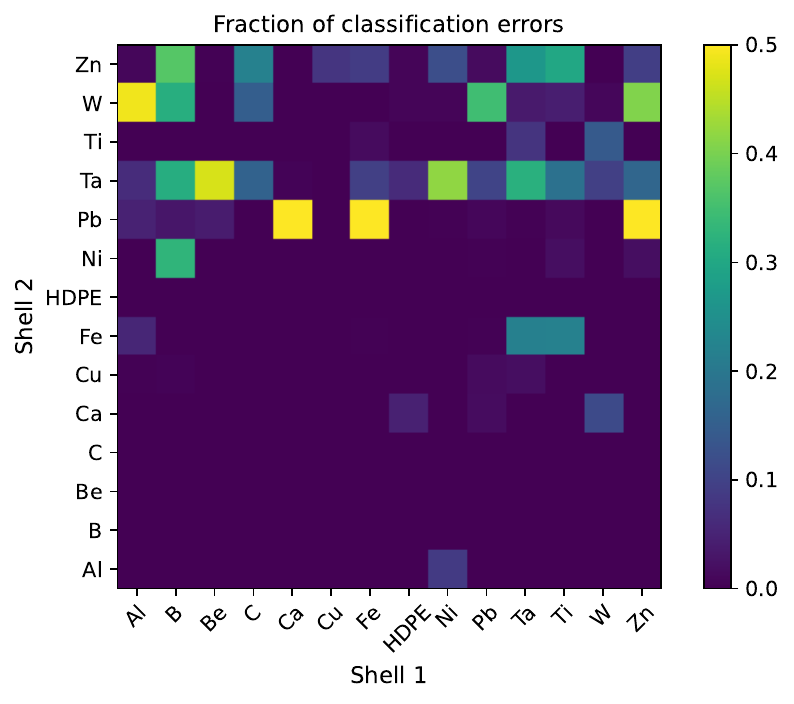}
  \caption{1.0/6.0, Gamma+$Y_2$}
  \label{fig:err_1p0_6p0_gammay2}
\end{subfigure}
\hfill
\begin{subfigure}[t]{0.35\linewidth}
  \centering
  \includegraphics[width=\linewidth]{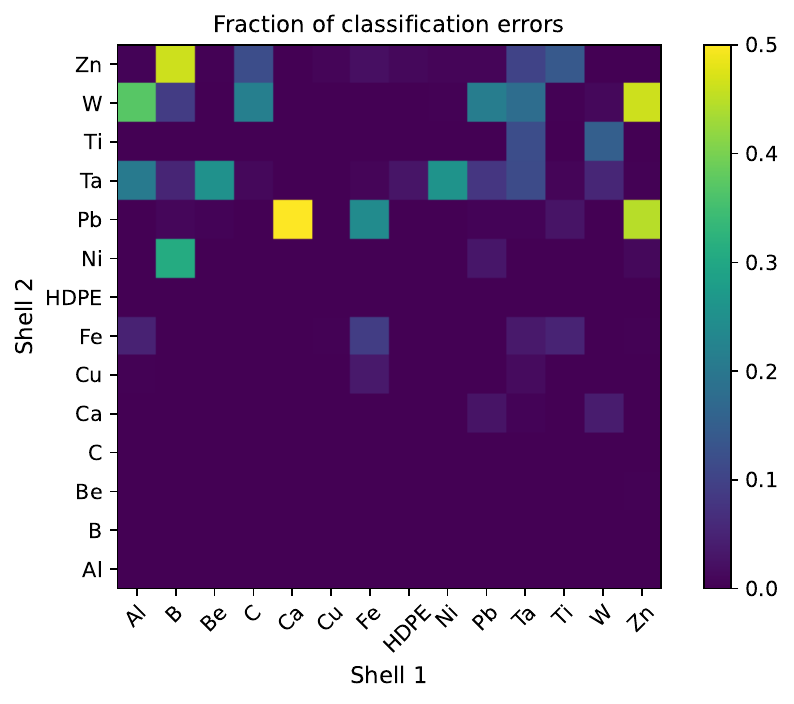}
  \caption{1.0/6.0, Gamma+$Y_2$+$Y_3$}
  \label{fig:err_1p0_6p0_gammay2y3}
\end{subfigure}
}
% \vspace{0.1cm}

% ---------------- Row 4: 6.0 / 1.0 ----------------
\resizebox{1\columnwidth}{!}{%
\begin{subfigure}[t]{0.35\linewidth}
  \centering
  \includegraphics[width=\linewidth]{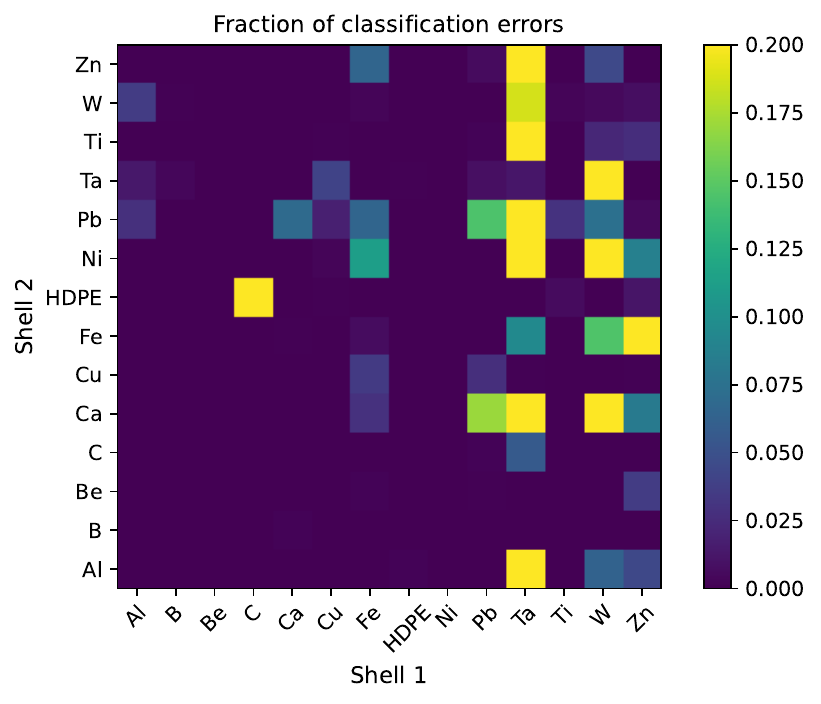}
  \caption{6.0/1.0, Gamma}
  \label{fig:err_6p0_1p0_gamma}
\end{subfigure}
\hfill
\begin{subfigure}[t]{0.35\linewidth}
  \centering
  \includegraphics[width=\linewidth]{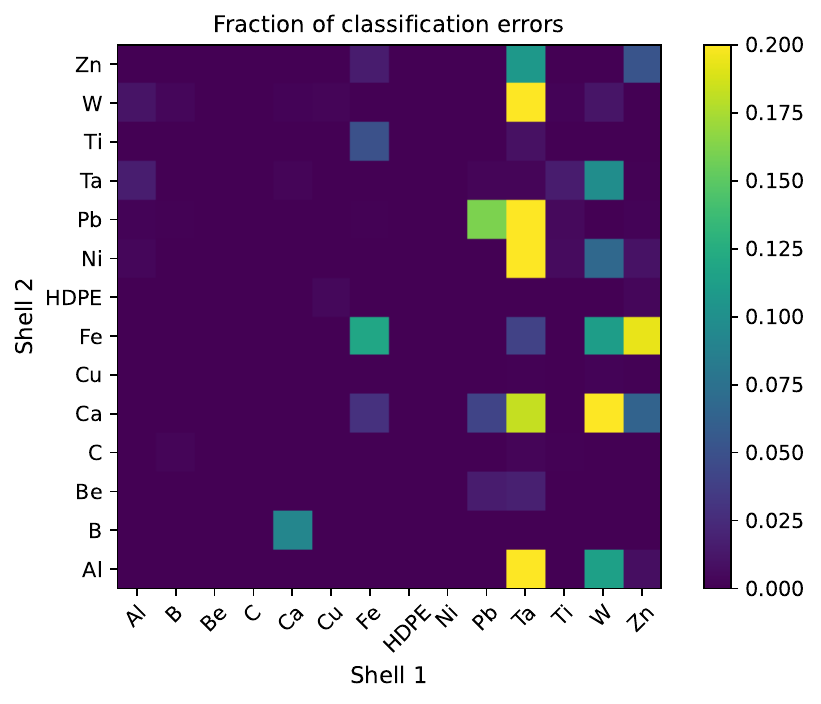}
  \caption{6.0/1.0, Gamma+$Y_2$}
  \label{fig:err_6p0_1p0_gammay2}
\end{subfigure}
\hfill
\begin{subfigure}[t]{0.35\linewidth}
  \centering
  \includegraphics[width=\linewidth]{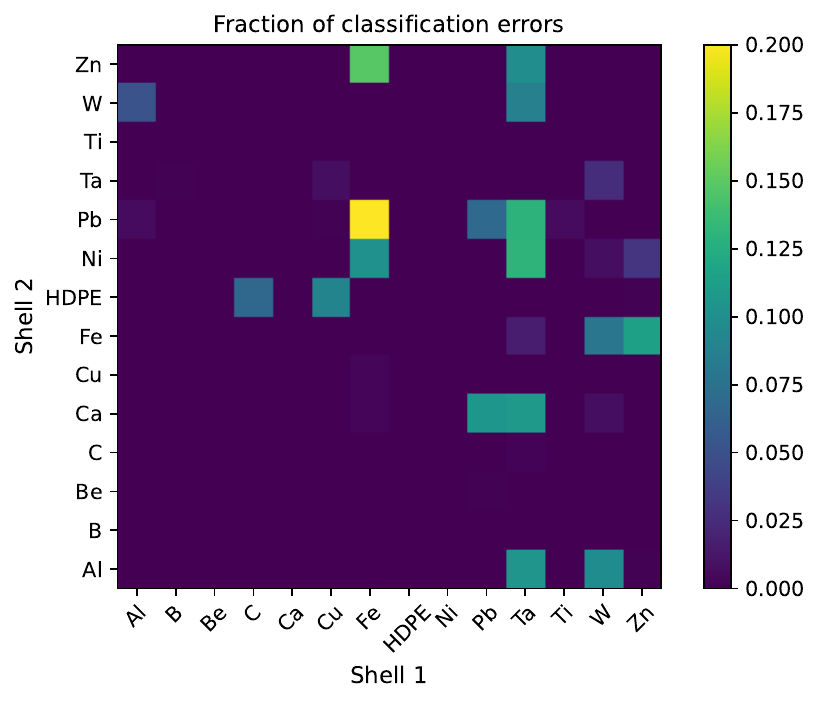}
  \caption{6.0/1.0, Gamma+$Y_2$+$Y_3$}
  \label{fig:err_6p0_1p0_gammay2y3}
\end{subfigure}
}

\caption{Misclassification fractions for representative two-shell configurations (rows) under three feature sets (columns): Gamma only (left), Gamma + $Y_2$ (middle), and Gamma + $Y_2$ + $Y_3$ (right) using 1800s time window and clean data for training. $N=10{,}000$ independent noise realizations per feature set and thickness are generated for testing. Each panel summarizes the fraction of test cases that are misclassified for the specified shell-thickness configuration (inner/outer).}
\label{fig:errors_2shell_symmetric}
\end{figure*}

\begin{figure*}[t]
\centering

% ---------------- Row 1: 0.5 / 6.0 ----------------
\resizebox{1\columnwidth}{!}{%
\begin{subfigure}[t]{0.35\linewidth}
  \centering
  \includegraphics[width=\linewidth]{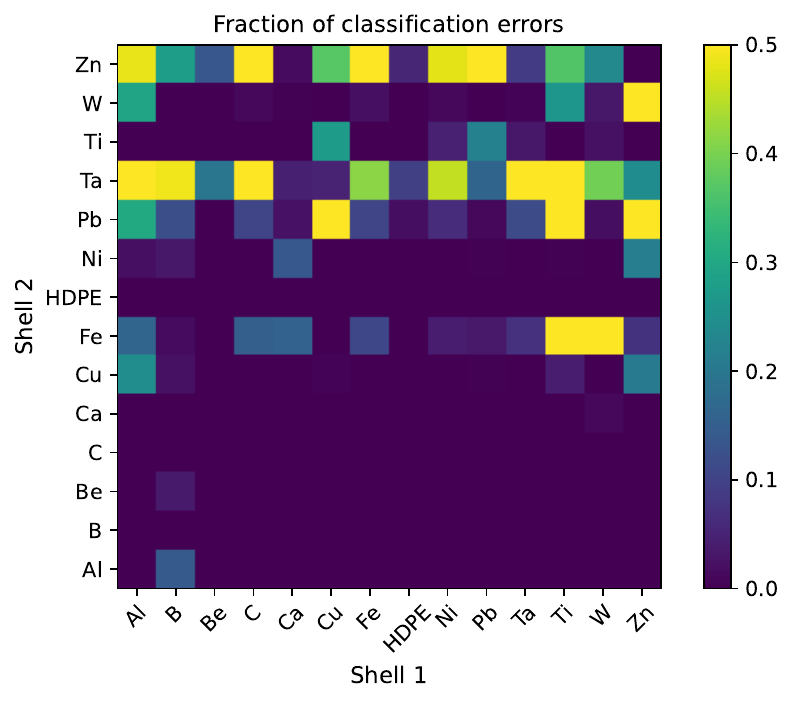}
  \caption{0.5/6.0, Gamma}
  \label{fig:err_0p5_6p0_gamma}
\end{subfigure}
\hfill
\begin{subfigure}[t]{0.35\linewidth}
  \centering
  \includegraphics[width=\linewidth]{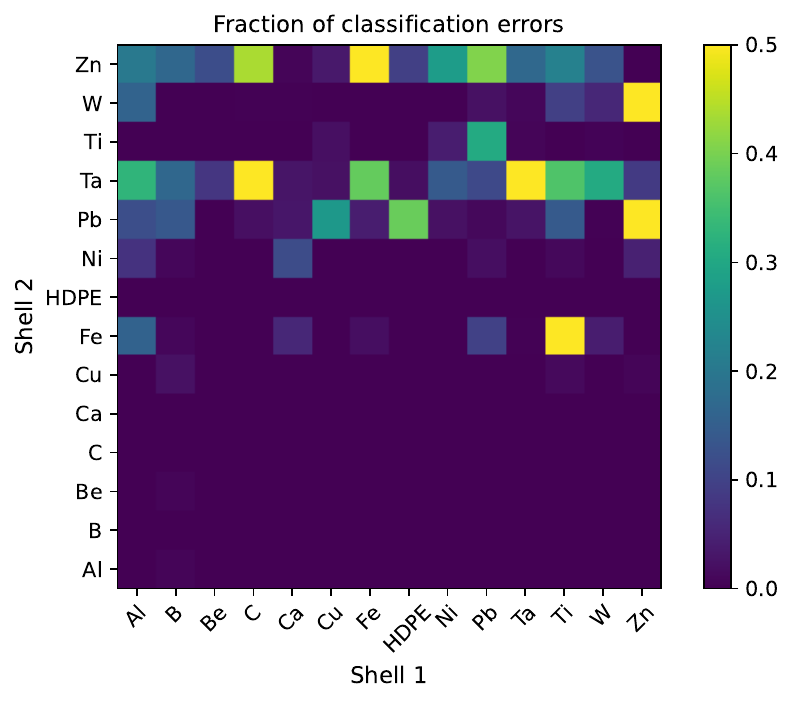}
  \caption{0.5/6.0, Gamma+$Y_2$}
  \label{fig:err_0p5_6p0_gammay2}
\end{subfigure}
\hfill
\begin{subfigure}[t]{0.35\linewidth}
  \centering
  \includegraphics[width=\linewidth]{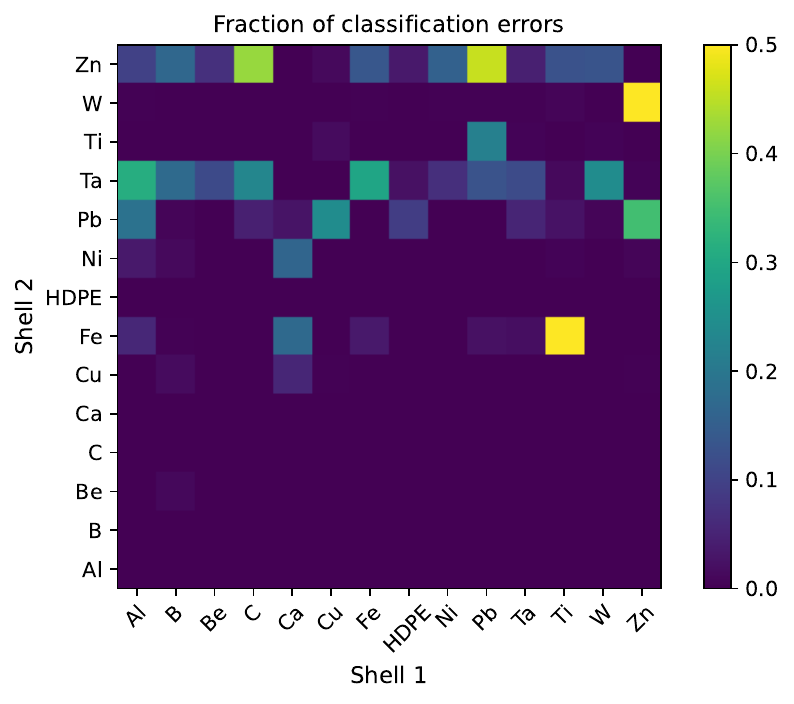}
  \caption{0.5/6.0, Gamma+$Y_2$+$Y_3$}
  \label{fig:err_0p5_6p0_gammay2y3}
\end{subfigure}
}
% \vspace{0.4cm}

% ---------------- Row 2: 6.0 / 3.0 ----------------
\resizebox{1\columnwidth}{!}{%
\begin{subfigure}[t]{0.35\linewidth}
  \centering
  \includegraphics[width=\linewidth]{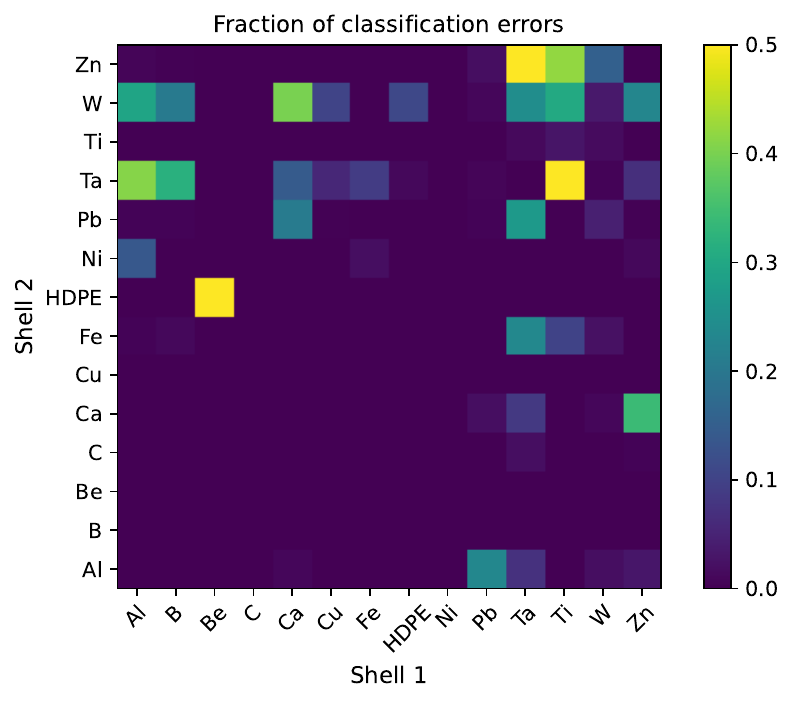}
  \caption{6.0/3.0, Gamma}
  \label{fig:err_6p0_3p0_gamma}
\end{subfigure}
\hfill
\begin{subfigure}[t]{0.35\linewidth}
  \centering
  \includegraphics[width=\linewidth]{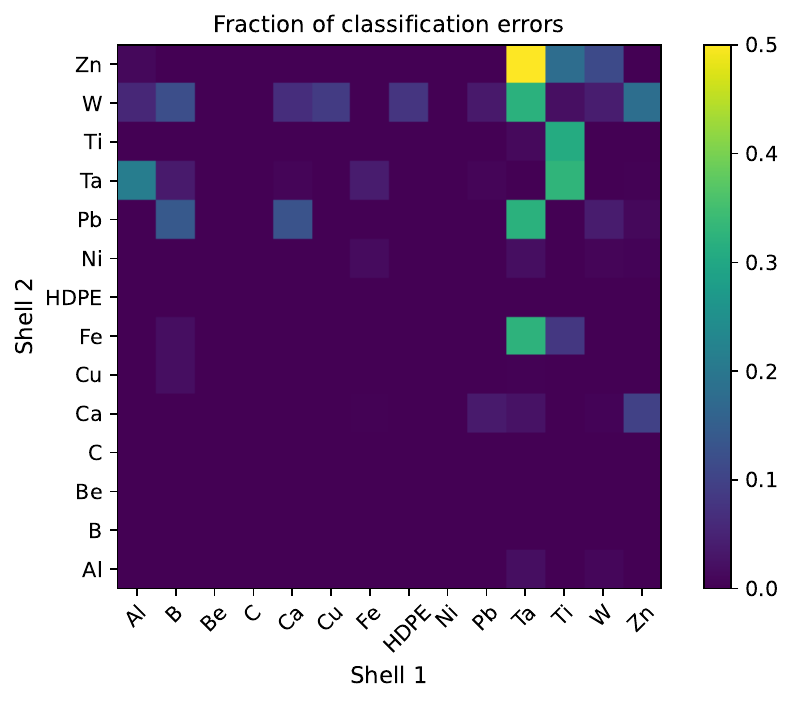}
  \caption{6.0/3.0, Gamma+$Y_2$}
  \label{fig:err_6p0_3p0_gammay2}
\end{subfigure}
\hfill
\begin{subfigure}[t]{0.35\linewidth}
  \centering
  \includegraphics[width=\linewidth]{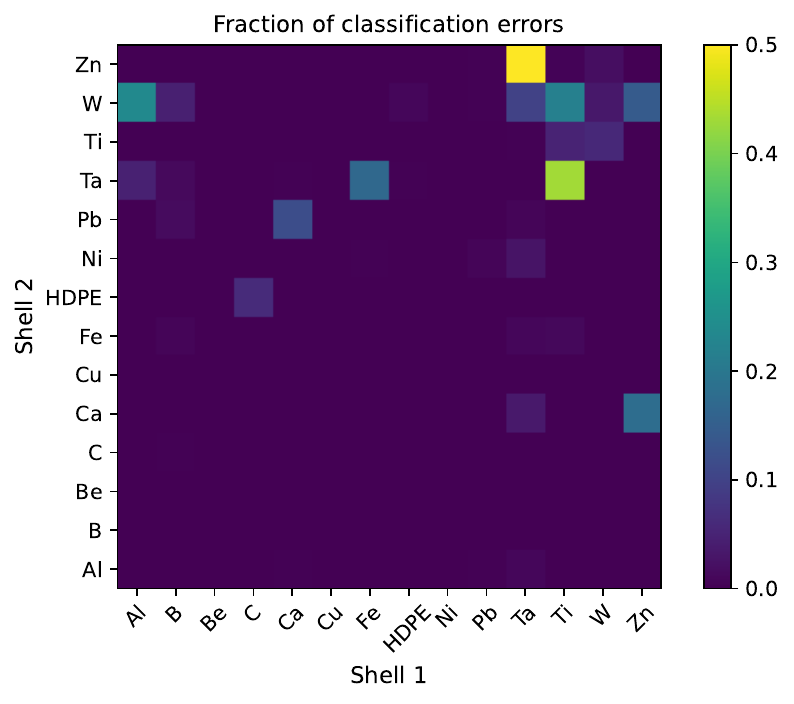}
  \caption{6.0/3.0, Gamma+$Y_2$+$Y_3$}
  \label{fig:err_6p0_3p0_gammay2y3}
\end{subfigure}
}
\caption{Misclassification fractions for asymmetric two-shell configurations (rows) under three feature sets (columns): Gamma only (left), Gamma + $Y_2$ (middle), and Gamma + $Y_2$ + $Y_3$ (right) using 1800s time window and clean data for training. $N=10{,}000$ independent noise realizations per feature set and thickness are generated for testing. The reduction in misclassification fraction when adding neutron multiplicity features highlights the complementary information provided by neutron transport relative to gamma attenuation in layered shielding.}
\label{fig:errors_2shell_asymmetric}
\end{figure*}

%In the gamma-only case, errors are broadly distributed across many material pairs, indicating diffuse ambiguity. With the inclusion of $Y_2$, misclassifications collapse onto a much smaller set of material pairs, primarily involving materials with similar neutron interaction properties. The addition of $Y_3$ produces only marginal further changes, consistent with its larger measurement uncertainty.
\begin{figure*}[t]
\centering

% ---------------- Row 1: 0.5 / 0.5 ----------------
\resizebox{1\columnwidth}{!}{%
\begin{subfigure}[t]{0.35\linewidth}
  \centering
  \includegraphics[width=\linewidth]{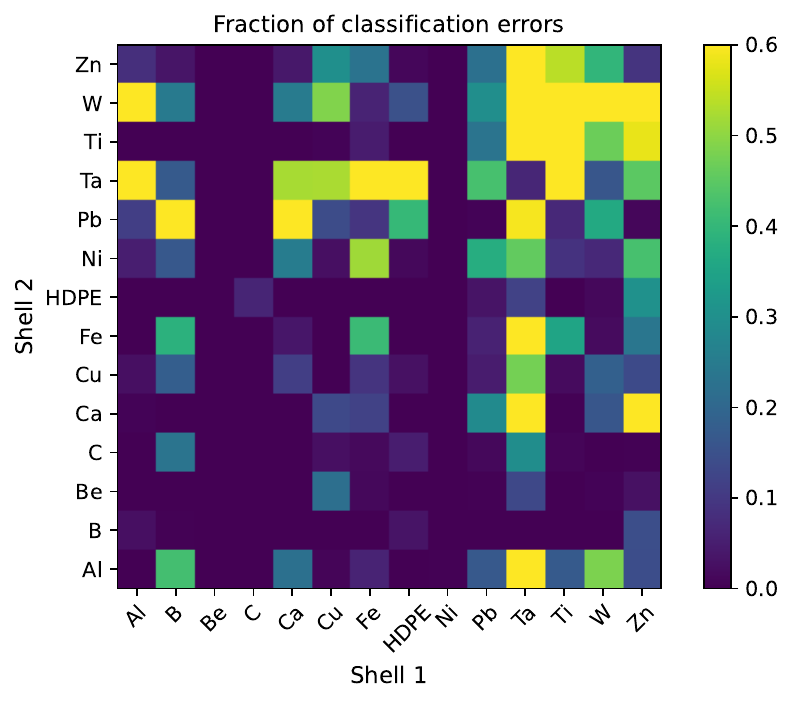}
  \caption{60s, clean data}
  \label{fig:err_6p0_3p0_60s_clean}
\end{subfigure}
\hfill
\begin{subfigure}[t]{0.35\linewidth}
  \centering
  \includegraphics[width=\linewidth]{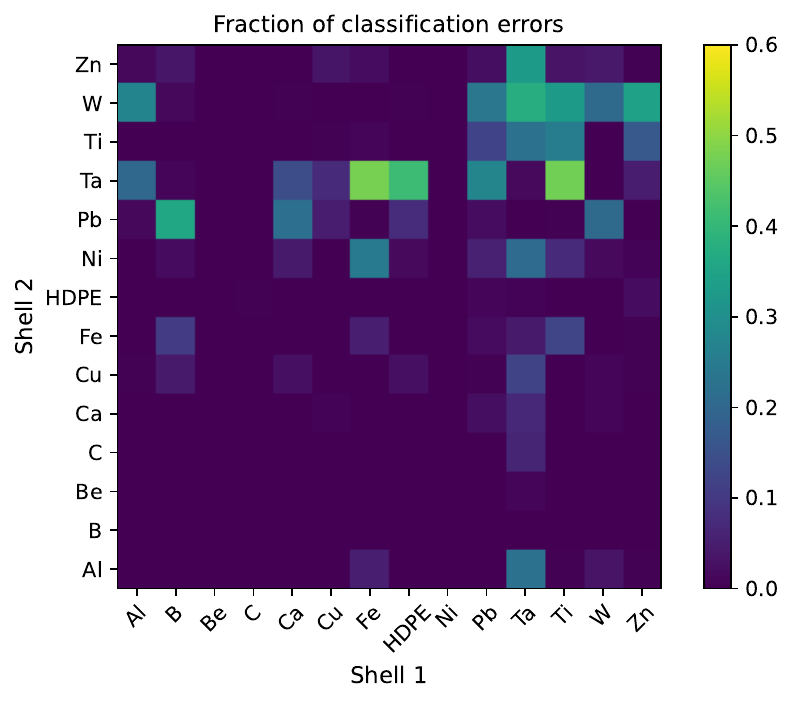}
  \caption{60s, 10 noisy samples/simulation}
  \label{fig:err_6p0_3p0_60s_10noise}
\end{subfigure}
\hfill
\begin{subfigure}[t]{0.35\linewidth}
  \centering
  \includegraphics[width=\linewidth]{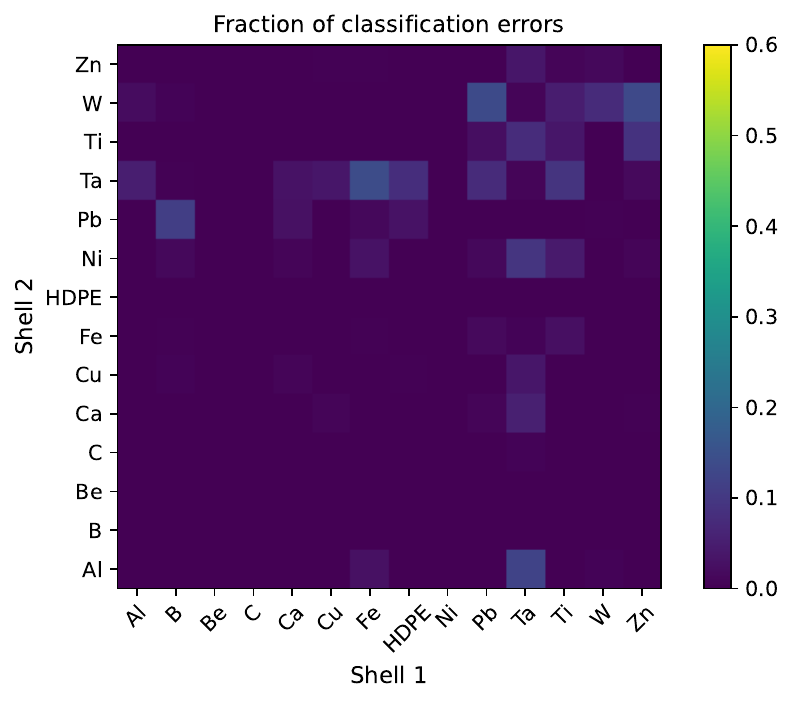}
  \caption{60s, 100 noisy samples/simulation}
  \label{fig:err_6p0_3p0_60s_100noise}
\end{subfigure}
}
% \vspace{0.4cm}

% ---------------- Row 2: 3.0 / 3.0 ----------------
\resizebox{1\columnwidth}{!}{%
\begin{subfigure}[t]{0.35\linewidth}
  \centering
  \includegraphics[width=\linewidth]{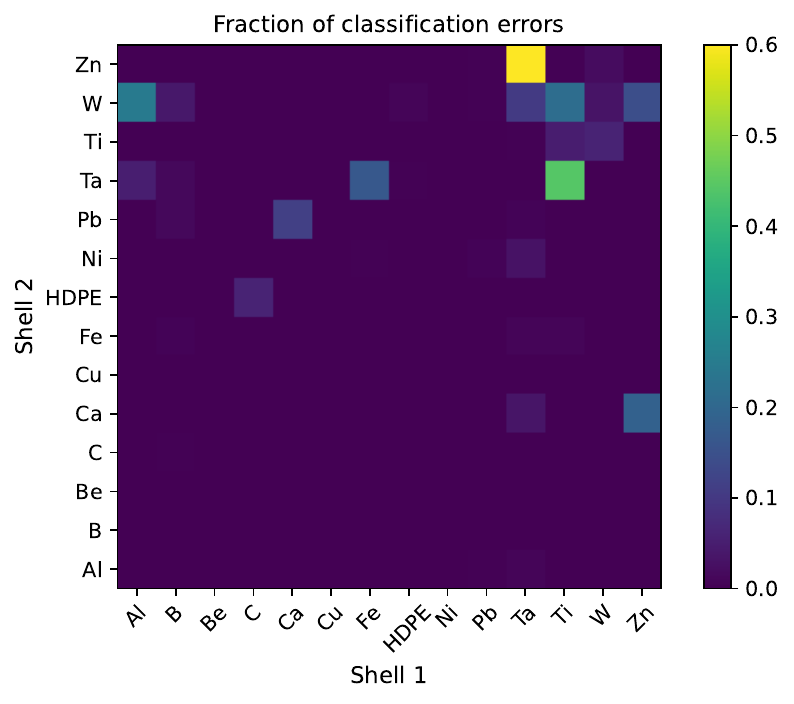}
  \caption{1800s, clean data}
  \label{fig:err_6p0_3p0_clean}
\end{subfigure}
\hfill
\begin{subfigure}[t]{0.35\linewidth}
  \centering
  \includegraphics[width=\linewidth]{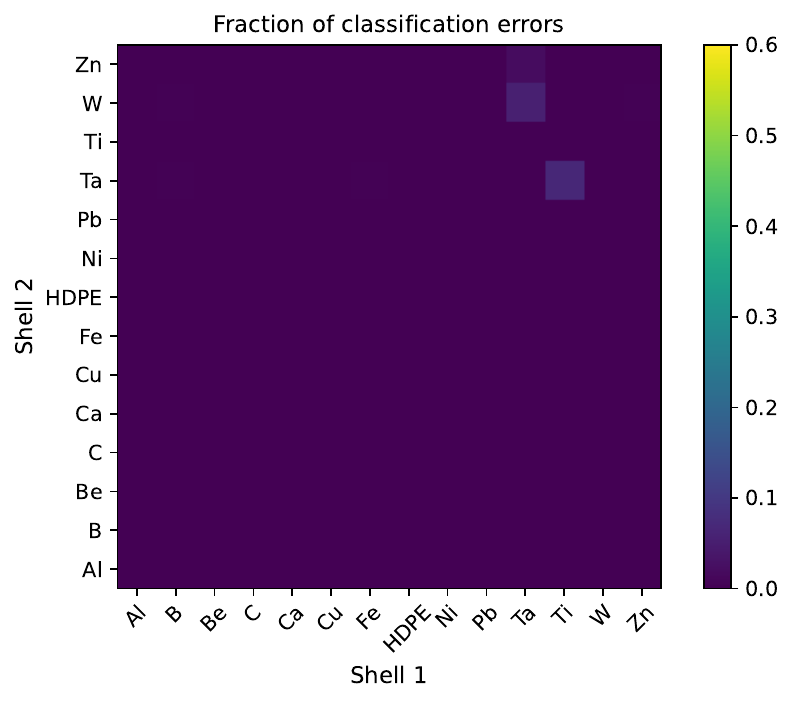}
  \caption{1800s, 10 noisy samples/simulation}
  \label{fig:err_6p0_3p0_10noise}
\end{subfigure}
\hfill
\begin{subfigure}[t]{0.35\linewidth}
  \centering
  \includegraphics[width=\linewidth]{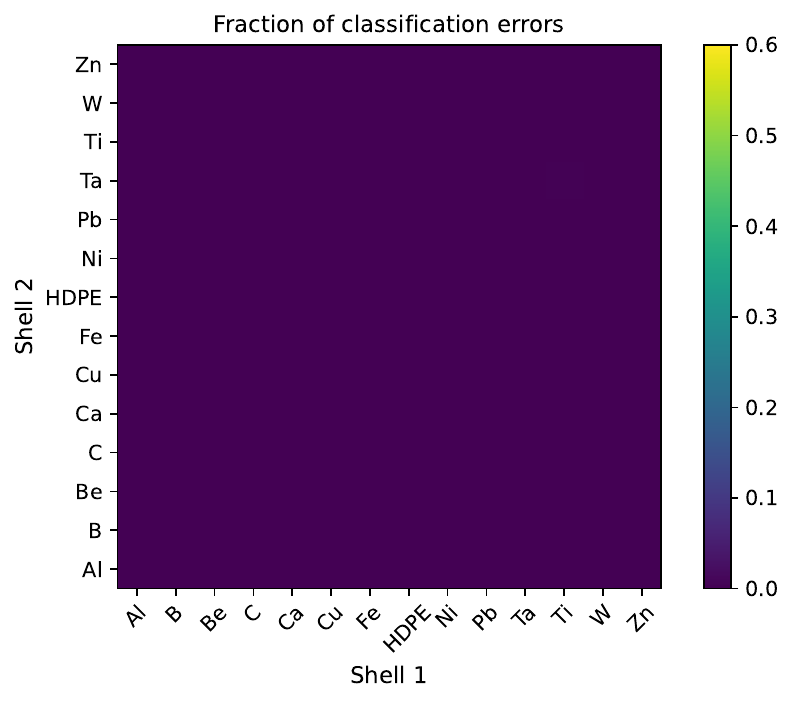}
  \caption{1800s, 100 noisy samples/simulation}
  \label{fig:err_6p0_3p0_100noise}
\end{subfigure}
}
\caption{Misclassification fractions for representative two-shell configurations (rows) under three configuration of training data (columns): clean data (left), 10 noisy samples per simulation (middle), and 100 noisy samples per simulation (right) using 6.0/3.0 in shells configuration, Gamma+$Y_2$+$Y_3$ as feature, and 60 or 1800s time window. $N=10{,}000$ independent noise realizations per feature set and thickness are generated for testing. Each panel summarizes the fraction of test cases that are misclassified for the specified shell-thickness and time window configuration (inner/outer).}
\label{fig:errors_2shell_symmetric_compare_window}
\end{figure*}

To quantify the structure of residual errors in the double-shell problem, we compute the entropy of the pairwise misclassification distribution using the error matrices reported in Figures \ref{fig:errors_2shell_symmetric} and \ref{fig:errors_2shell_asymmetric}.  The results of this evaluation are presented in Table \ref{tab:double_shell_entropy}. For the 0.5~in / 0.5~in configuration, the gamma-only case exhibits a high entropy of $H=3.34$~nats, corresponding to approximately 28 effective error modes, indicating diffuse misclassification across many material-pair combinations.

Incorporating the neutron multiplicity feature $Y_2$ reduces the entropy to $H=2.62$~nats, collapsing the effective number of error modes to approximately 14. This represents a reduction of roughly 50\% in the number of dominant misclassification pathways, demonstrating that neutron multiplicity information does not merely reduce overall error rate, but fundamentally restructures the error distribution by resolving gamma-degenerate material pairs.

The importance in characterizing the noise or minimizing the mismatch between the forward operators and data is examined in Figure \ref{fig:errors_2shell_symmetric_compare_window}. When the training data is augmented with noise samples with long count times it is observed that almost perfect classification is performed. Consequently, in the limit of large data augmentation, or no model mismatch, near perfect double-shell classification can be performed using the multi-modality algorithm.

To aid interpretation of the per-shell confusion matrices, we also decompose the
pairwise classification errors, Figure \ref{fig:double_shell_error_shell1_2_both}, into cases where (i) only shell~1 is misidentified,
(ii) only shell~2 is misidentified, or (iii) both shells are misidentified.
This error attribution clarifies whether residual ambiguity in a given
configuration arises predominantly from the inner or outer layer, and it
separates isolated shell errors from coupled misclassifications.

Figure \ref{fig:errors_2shell_confusion} presents per-shell confusion matrices for the double-shell problem
using the full multi-modal feature set ($\gamma + Y_2 + Y_3$). For each panel,
rows correspond to the true material in the indicated shell, columns correspond
to the predicted material for that shell, and each row is normalized to unity.
The influence of the other shell is marginalized, so the matrices quantify how
often a given material is misclassified independent of the composition of the
other layer.

As an illustration of the application of how Figures \ref{fig:double_shell_error_shell1_2_both} and \ref{fig:errors_2shell_confusion} may be utilized to interpret the double shell configuration Zn/W in Figure \ref{fig:errors_2shell_symmetric} from which we see that approximately 15 percent are misclassified.  Examination of Figure \ref{fig:double_shell_error_shell1_2_both} indicates that all of these misidentified cases are attributed to shell one.  Finally, examination of \ref{fig:errors_2shell_confusion}
reveals that the misclassification is attributed to the confusion of Zn with W in the first shell.

\

\begin{figure*}[t]
    \centering % Centers the figure
    \resizebox{1\columnwidth}{!}{%
\begin{subfigure}[t]{0.35\linewidth}
  \centering
  \includegraphics[width=\linewidth]{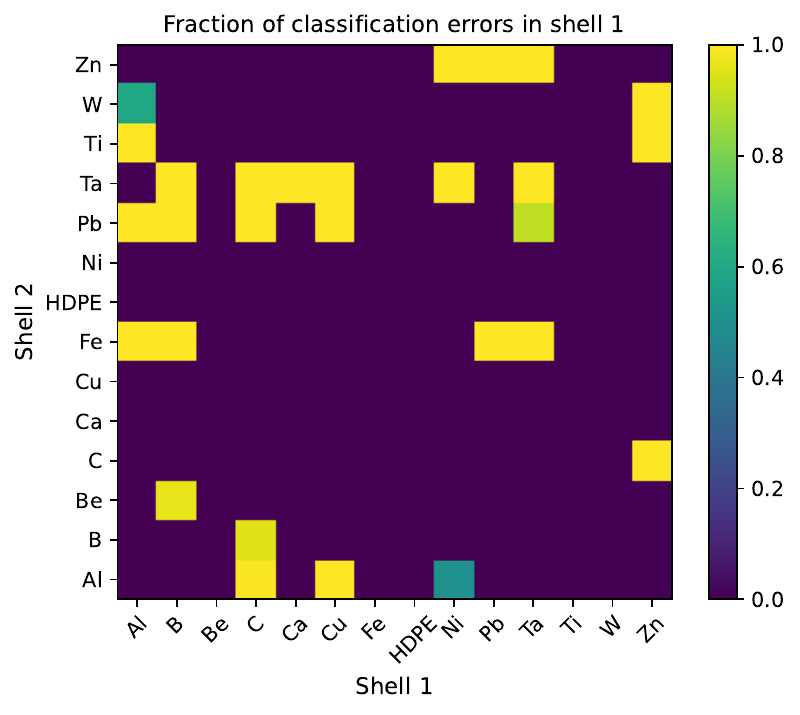}
  \caption{Shell 1 only}
  \label{fig:err_6p0_3p0_shell1}
\end{subfigure}
\begin{subfigure}[t]{0.35\linewidth}
  \centering
  \includegraphics[width=\linewidth]{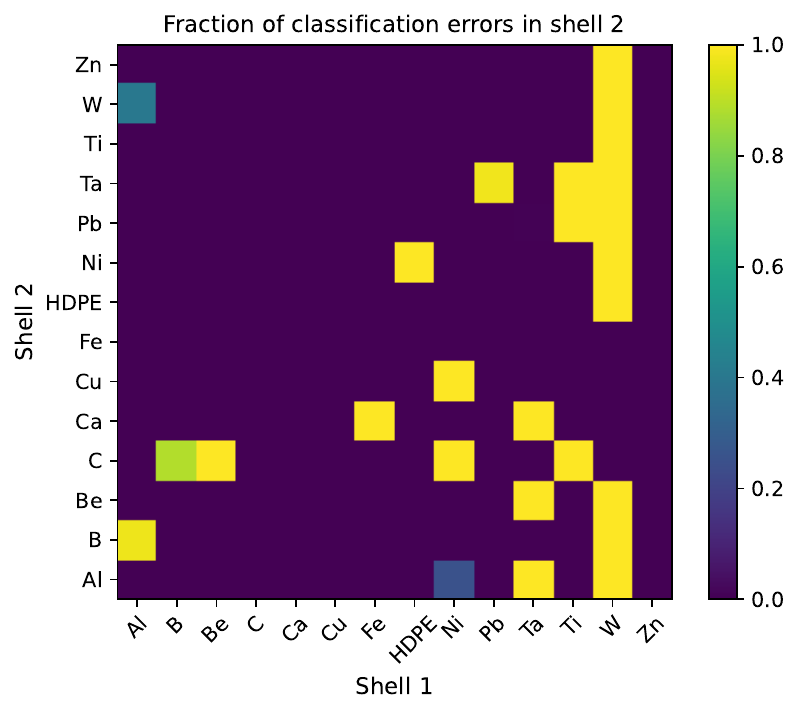}
  \caption{Shell 2 only}
  \label{fig:err_6p0_3p0_shell2}
\end{subfigure}
\begin{subfigure}[t]{0.35\linewidth}
  \centering
  \includegraphics[width=\linewidth]{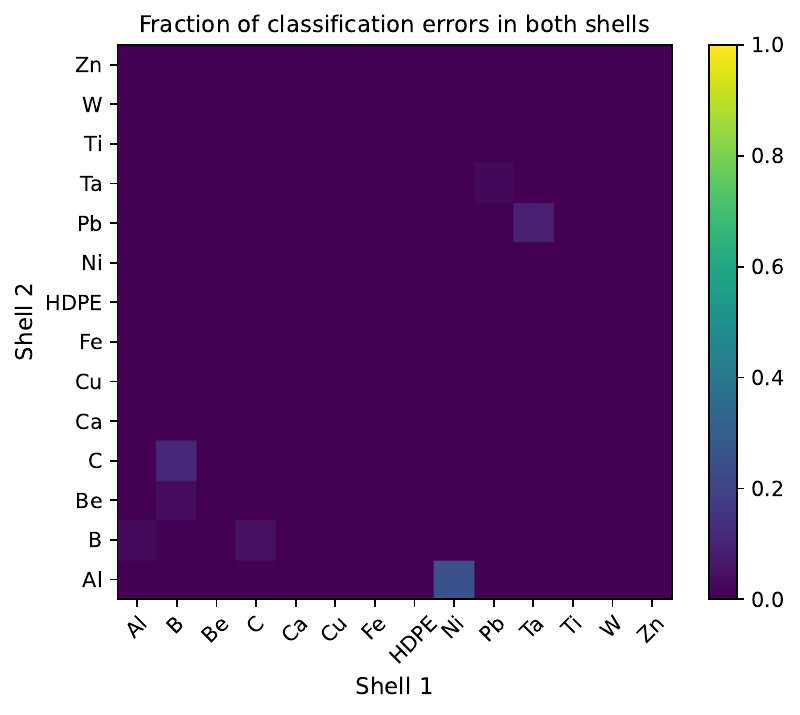}
  \caption{Both shells}
  \label{fig:err_6p0_3p0_shell12}
\end{subfigure}
} 
    \caption{Fraction of classification errors for (a) only shell 1 is misidentified, (b) only shell 2 is misidentified, and (c) both shells are misidentified, by material and thickness in double shell configuration with 0.5/0.5 thickness, using 1800s time window and Gamma+$Y_2$+$Y_3$ as feature. We normalize these errors so that for each material pair, the sum of errors in (a), (b), and (c) is $1$ (if they are non-zeros). } 
    \label{fig:double_shell_error_shell1_2_both} % 
\end{figure*}

\begin{figure*}[t]
\centering

% ---------------- Row 1: 0.5 / 0.5 ----------------
\resizebox{0.9\columnwidth}{!}{%
\begin{subfigure}[t]{0.45\linewidth}
  \centering
  \includegraphics[width=\linewidth]{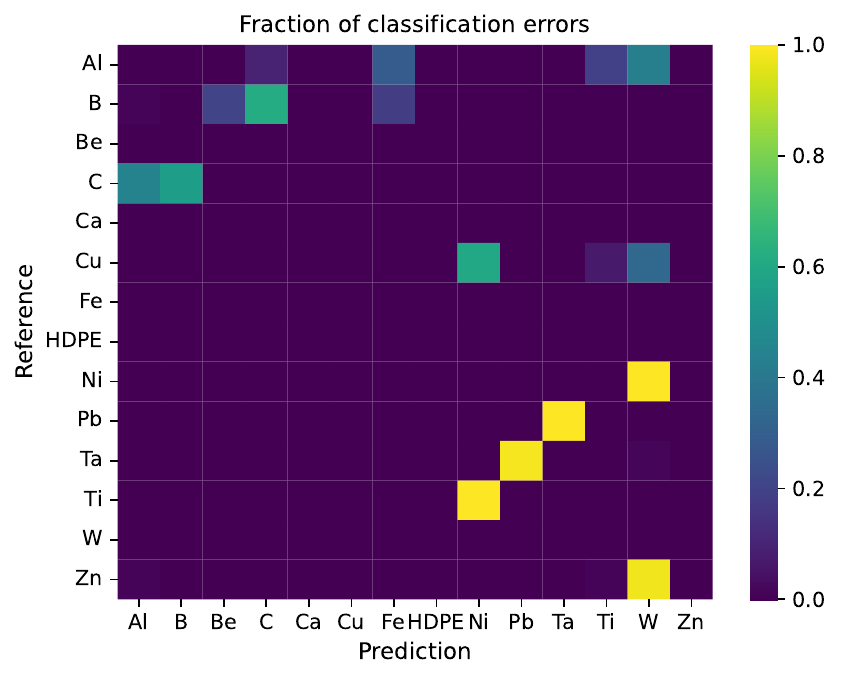}
  \caption{0.5/0.5, Shell 1}
  \label{fig:err_0p5_0p5_confusion_shell0}
\end{subfigure}
% \hfill
\begin{subfigure}[t]{0.45\linewidth}
  \centering
  \includegraphics[width=\linewidth]{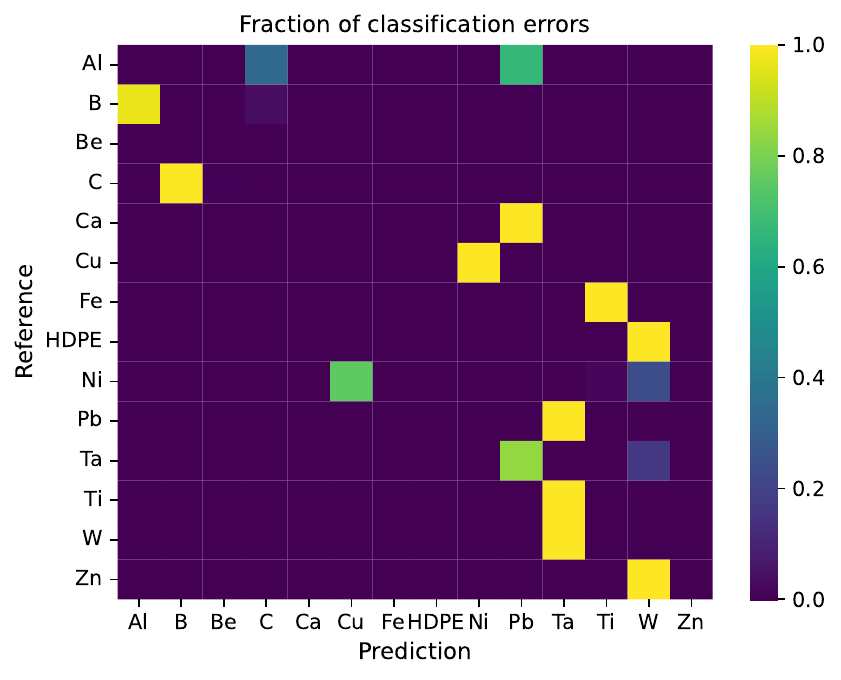}
  \caption{0.5/0.5, Shell 2}
  \label{fig:err_0p5_0p5_confusion_shell1}
\end{subfigure}
}
% \vspace{0.4cm}

% ---------------- Row 2: 1.0 / 6.0 ----------------
\resizebox{0.9\columnwidth}{!}{%
\begin{subfigure}[t]{0.45\linewidth}
  \centering
  \includegraphics[width=\linewidth]{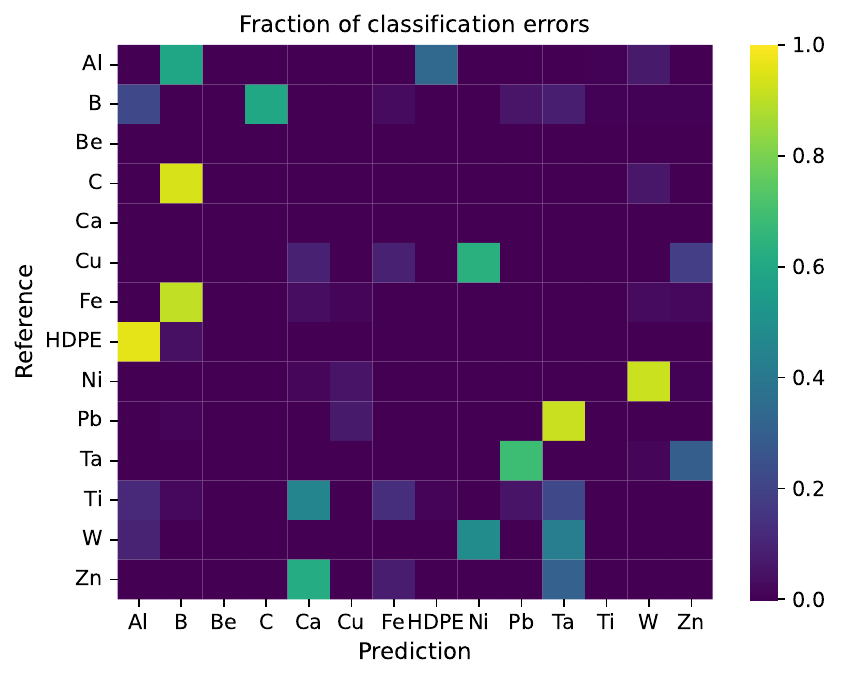}
  \caption{1.0/6.0, Shell 1}
  \label{fig:err_1p0_6p0_confusion_shell0}
\end{subfigure}
% \hfill
\begin{subfigure}[t]{0.45\linewidth}
  \centering
  \includegraphics[width=\linewidth]{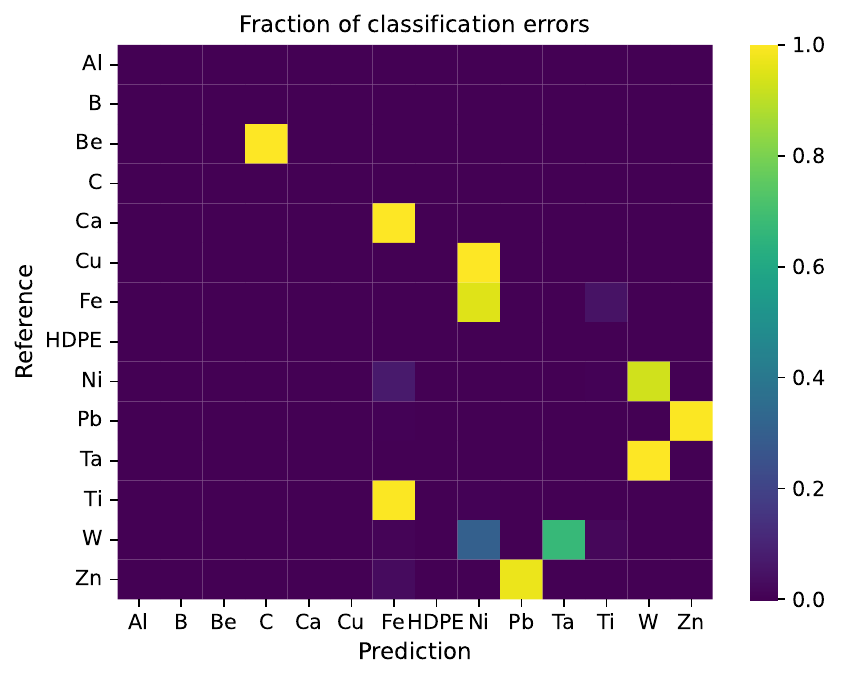}
  \caption{1.0/6.0, Shell 2}
  \label{fig:err_1p0_6p0_confusion_shell1}
\end{subfigure}
}

\caption{Normalized confusion metric with 0.5/0.5 shell thickness using 1800s time window, Gamma+$Y_2$+$Y_3$ as feature, and clean data for training. $N=10{,}000$ independent noise realizations per feature set and thickness are generated for testing. We normalize these matrices so that the sum of classification errors for each reference material is 1 (if they are non-zeros).}
\label{fig:errors_2shell_confusion}
\end{figure*}

Taken together, the two-shell results demonstrate that (i) multi-modality is essential once multiple shielding layers are present, (ii) neutron multiplicity features substantially reduce ordering-induced ambiguities inherent to gamma-only classification, and (iii) longer acquisitions and sufficiently rich noise augmentation improve robustness, with performance saturating once the noise distribution is adequately represented in training.
\section{Conclusions}
\label{sec:conclusions}

This work examined multi-modality material identification for spherically symmetric shielding configurations surrounding a BeRP plutonium metal core, using radiography-derived shell thicknesses together with HPGe $\gamma$-ray photopeak counts and neutron multiplicity features summarized by the Feynman variances $Y_2$ and $Y_3$. For single-shell configurations, classification accuracy is high across the thicknesses examined using gamma spectroscopy alone, with the most pronounced residual confusions occurring in regimes where either (i) the shell is thin and produces limited contrast or (ii) the shell is sufficiently thick that the usable spectral information is strongly attenuated. In these regimes, the addition of neutron multiplicity---particularly $Y_2$---consistently reduces residual confusion by introducing sensitivity to moderation and absorption properties that are complementary to gamma attenuation.

Two additional observations are noteworthy. (i) The confusion structure is asymmetric across the two shell indices: misclassification patterns for ``inner'' and ``outer'' shell labels differ because gamma attenuation is inherently directional — when a strongly attenuating layer is nearer the source it disproportionately modifies the spectrum seen by the detector. Neutron multiplicity (captured primarily by $Y_2$) mitigates this asymmetry because $Y_2$ depends on the integrated moderation/absorption behavior of the entire assembly rather than on radial ordering alone, but it does not eliminate ordering-dependent residuals. (ii) The entropy diagnostic reported in Table~\ref{tab:double_shell_entropy} quantitatively supports the visual impressions: including neutron multiplicity reduces the entropy of the misclassification distribution (fewer effective error modes), demonstrating that $Y_2$ changes the geometry of class overlap rather than simply reducing variance uniformly. 

Taken together, these results show that (a) gamma spectroscopy alone is insufficient for reliable two-shell identification in many realistic thin–thin and asymmetric configurations, (b) $Y_2$ is the dominant multiplicity feature that collapses large-scale ambiguity into a small set of physically-meaningful confusions, and (c) higher-order multiplicity moments (e.g., $Y_3$) offer only incremental value at the noise levels considered here. We therefore recommend experimental follow-up focused on the small set of persistently confused material pairs (particularly those involving light moderators and boron-containing compounds) and on improved characterization of $Y_3$ uncertainty before ascribing operational value to third-order multiplicity features.
% ===================== END REPLACE =====================

These results demonstrate that neutron multiplicity provides essential complementary information for multi-layer shielding identification and motivates extension of the proposed framework to more complex geometries, broader material sets, and experimental measurements where model mismatch and environmental background must be explicitly addressed.

\section{Future Work}
\label{sec:future_work}

The present study demonstrates the potential of multi-modal material identification using gamma spectroscopy and neutron multiplicity for idealized, spherically symmetric shielding configurations with known layer interfaces. Several important directions for future investigation remain.

A primary extension is to general three-dimensional (3D) objects with arbitrary geometries. In realistic inspection scenarios, shielding configurations are unlikely to be concentric or spherically symmetric, and interfaces between materials may be irregular or only partially resolved by radiography. Extending the proposed framework to 3D geometries will require coupling multi-view or tomographic radiographic reconstructions with localized gamma and neutron feature extraction, as well as adapting the classification framework to operate on spatially heterogeneous feature sets. The ability to handle partial-volume effects, segmentation uncertainty, and unknown interface topology will be critical for practical deployment.

Experimental validation represents another essential next step. While the present results are based on high-fidelity GADRAS simulations with physics-informed noise models, real measurements introduce additional complexities, including background radiation, detector dead time, environmental scattering, and imperfect detector calibration. Previously experimental validation was performed using gamma spectroscopic measurements, controlled multi-modal experiments using well-characterized plutonium or surrogate sources with known shielding compositions and thicknesses would enable direct validation of the multi-modality approach and provide empirical bounds on achievable accuracy under realistic operating conditions. Such experiments would also help refine uncertainty models for neutron multiplicity observables, particularly for higher-order moments such as $Y_3$.

Further work is also warranted to address uncertainty in source characteristics. In this study, the SNM isotopic composition and mass were assumed known, consistent with scenarios in which isotope identification precedes shielding analysis. In practice, uncertainties in isotopics, multiplication, or $(\alpha,n)$ contributions may be present. Joint inference frameworks that simultaneously estimate source properties and shielding composition, or Bayesian approaches that marginalize over plausible source parameters, could improve robustness in these settings.

Finally, extensions to alternative feature representations and learning paradigms merit investigation. While random forests performed well in the present study, hierarchical models, structured classifiers, or hybrid physics–machine-learning approaches may better exploit known physical constraints, such as ordering effects and monotonic attenuation behavior. Incorporating temporal information, additional neutron observables, or complementary sensing modalities may further enhance performance in complex shielding scenarios.

Together, these directions point toward a comprehensive, experimentally validated multi-modal framework for material identification in realistic nuclear security and safeguards applications.

\section{Acknowledgments}
The authors would like to acknowledge the funding for this research provided by the National Nuclear Security Administration Office-NA-22 and the resources provided by the Los Alamos National Laboratory.

\bibliography{mybibfile}

@book{breiman1984classification,
  title={Classification and Regression Trees},
  author={Breiman, Leo and Friedman, Jerome H. and Stone, Charles J. and Olshen, Richard A.},
  isbn={9780412048418},
  year={1984},
  publisher={Taylor \& Francis}
}

@article{scikit-learn,
  title =	 {Scikit-learn: Machine Learning in {P}ython},
  author =	 {Pedregosa, F. and Varoquaux, G. and Gramfort, A. and
                  Michel, V.  and Thirion, B. and Grisel, O. and
                  Blondel, M. and Prettenhofer, P.  and Weiss, R. and
                  Dubourg, V. and Vanderplas, J. and Passos, A. and
                  Cournapeau, D. and Brucher, M. and Perrot, M. and
                  Duchesnay, E.},
  journal =	 {Journal of Machine Learning Research},
  volume =	 12,
  pages =	 {2825--2830},
  year =	 2011
}

@article{alvarez1976,
doi = {10.1088/0031-9155/21/5/002},
IGNOREurl = {https://dx.doi.org/10.1088/0031-9155/21/5/002},
year = {1976},
month = {sep},
publisher = {},
volume = {21},
number = {5},
pages = {733},
author = {R E Alvarez and  A Macovski},
title = {Energy-selective reconstructions in {X}-ray computerised tomography},
journal = {Physics in Medicine \& Biology}
}

@Article{beldjoudi2012,
  author   = {Guillaume Beldjoudi and Véronique Rebuffel and Loïck Verger and Valérie Kaftandjian and Jean Rinkel},
  journal  = {Nuclear Instruments and Methods in Physics Research Section A: Accelerators, Spectrometers, Detectors and Associated Equipment},
  title    = {An optimised method for material identification using a photon counting detector},
  year     = {2012},
  issn     = {0168-9002},
  number   = {1},
  pages    = {26-36},
  volume   = {663},
  doi      = {10.1016/j.nima.2011.09.002},
  IGNOREurl      = {https://www.sciencedirect.com/science/article/pii/S0168900211017451},
}

@article{cann1982,
  title={Quantification of calcium in solitary pulmonary nodules using single-and dual-energy CT.},
  author={Cann, Christopher E and Gamsu, G and Birnberg, FA and Webb, WR},
  journal={Radiology},
  volume={145},
  number={2},
  pages={493--496},
  year={1982}
}

@article{fraser1986,
  title={Potential value of digital radiography: preliminary observations on the use of dual-energy subtraction in the evaluation of pulmonary nodules},
  author={Fraser, Robert G and Barnes, Gary T and Hickey, Nancy and Luna, Rodrigo and Katzenstein, Anna and Alexander, Bruce and McElvein, Richard and Zorn, George and Sabbagh, Eduardo and Robinson Jr, CA},
  journal={Chest},
  volume={89},
  number={4},
  pages={249S--252S},
  year={1986},
  publisher={Elsevier}
}

@book{herman2007,
  title     = "Advances in Discrete Tomography and Its Applications",
  editor    = "{Herman}, Gabor~T. and {Kuba}, Attila",
  year      = 2007,
  publisher = "Birkh{\"a}user",
  address   = "Boston - Basel - Berlin"
}

@article{kalender1986,
  title={Evaluation of a prototype dual-energy computed tomographic apparatus. I. Phantom studies},
  author={Kalender, Willi A and Perman, WH and Vetter, JR and Klotz, Ernst},
  journal={Medical physics},
  volume={13},
  number={3},
  pages={334--339},
  year={1986},
  publisher={Wiley Online Library}
}

@article{kelcz1979,
  title={Noise considerations in dual energy {CT} scanning},
  author={Kelcz, Frederick and Joseph, Peter M and Hilal, Sadek K},
  journal={Medical physics},
  volume={6},
  number={5},
  pages={418--425},
  year={1979},
  publisher={Wiley Online Library}
}

@ARTICLE{khatiwada23,
       author = {{Khatiwada}, Ajeeta and {Klasky}, Marc and {Lombardi}, Marcie and {Matheny}, Jason and {Mohan}, Arvind},
        title = "{Machine Learning technique for isotopic determination of radioisotopes using HPGe {\ensuremath{\gamma}}-ray spectra}",
      journal = {Nuclear Instruments and Methods in Physics Research A},
         year = 2023,
        month = sep,
       volume = {1054},
          eid = {168409},
        pages = {168409},
          doi = {10.1016/j.nima.2023.168409},
archivePrefix = {arXiv},
       eprint = {2301.01415},
 primaryClass = {physics.data-an},
       adsurl = {https://ui.adsabs.harvard.edu/abs/2023NIMPA105468409K}
}

@book{knoll2010,
  title     = "Radiation Detection and Measurement",
  editor    = "{Knoll}, Glenn~F.",
  year      = 2010,
  month     = aug,
  publisher = "Wiley",
  isbn      = "978-0470131480",
  edition   = "4."
}

@ARTICLE{long2014,
  author={Long, Yong and Fessler, Jeffrey A.},
  journal={IEEE Transactions on Medical Imaging}, 
  title={Multi-Material Decomposition Using Statistical Image Reconstruction for Spectral CT}, 
  year={2014},
  volume={33},
  number={8},
  pages={1614-1626},
  doi={10.1109/TMI.2014.2320284}
}

@article{mccann2023,
  title={Material Identification From Radiographs Without Energy Resolution},
  author={McCann, Michael T and Guardincerri, Elena and Gonzales, Samuel M and Misurek, Lauren A and Schei, Jennifer L and Klasky, Marc L},
  journal={IEEE Transactions on Computational Imaging},
  volume={9},
  pages={314--326},
  year={2023},
  publisher={IEEE}
}

@inproceedings{mendoncca2010,
  title={Multi-material decomposition of spectral {CT} images},
  author={Mendon{\c{c}}a, Paulo RS and Bhotika, Rahul and Maddah, Mahnaz and Thomsen, Brian and Dutta, Sandeep and Licato, Paul E and Joshi, Mukta C},
  booktitle={Medical Imaging 2010: Physics of Medical Imaging},
  volume={7622},
  pages={633--641},
  year={2010},
  organization={SPIE}
}

@article{mendoncca2013,
  title={A flexible method for multi-material decomposition of dual-energy {CT} images},
  author={Mendon{\c{c}}a, Paulo RS and Lamb, Peter and Sahani, Dushyant V},
  journal={IEEE transactions on medical imaging},
  volume={33},
  number={1},
  pages={99--116},
  year={2013},
  publisher={IEEE}
}

@misc{mitchell2009,
title = {Gamma Detector Response and Analysis Software (GADRAS) v. 16.0, Version 01},
author = {Mitchell, Dean and Mattingly, John},
url = {https://www.osti.gov/biblio/1231259}, 
year = {2009},
month = {12},
}

@article{naydenov2004,
  title={Direct reconstruction of the effective atomic number of materials by the method of multi-energy radiography},
  author={Naydenov, Sergey V and Ryzhikov, Vladimir D and Smith, Craig F},
  journal={Nuclear Instruments and Methods in Physics Research Section B: Beam Interactions with Materials and Atoms},
  volume={215},
  number={3-4},
  pages={552--560},
  year={2004},
  publisher={Elsevier}
}

@article{ogorodnikov2002,
  title={Processing of interlaced images in 4--10 MeV dual energy customs system for material recognition},
  author={Ogorodnikov, S and Petrunin, V},
  journal={Physical Review Special Topics-Accelerators and Beams},
  volume={5},
  number={10},
  pages={104701},
  year={2002},
  publisher={APS}
}

@article{paziresh2016,
  title={Tomography of atomic number and density of materials using dual-energy imaging and the Alvarez and Macovski attenuation model},
  author={Paziresh, M and Kingston, AM and Latham, SJ and Fullagar, WK and Myers, GM},
  journal={Journal of Applied Physics},
  volume={119},
  number={21},
  year={2016},
  publisher={AIP Publishing}
}

@article{runkle2009,
  title={Photon and neutron interrogation techniques for chemical explosives detection in air cargo: A critical review},
  author={Runkle, Robert C and White, Timothy A and Miller, Erin A and Caggiano, Joseph A and Collins, Brian A},
  journal={Nuclear Instruments and Methods in Physics Research Section A: Accelerators, Spectrometers, Detectors and Associated Equipment},
  volume={603},
  number={3},
  pages={510--528},
  year={2009},
  publisher={Elsevier}
}

@ARTICLE{tobias2019,
       author = {{Tobias}, B. and {Klasky}, M.~L.},
        title = "{Limits of special material detectability fundamental to idealized dual-energy radiographic systems}",
      journal = {Nuclear Instruments and Methods in Physics Research A},
         year = 2019,
        month = nov,
       volume = {944},
          eid = {162563},
        pages = {162563},
          doi = {10.1016/j.nima.2019.162563},
       adsurl = {https://ui.adsabs.harvard.edu/abs/2019NIMPA.94462563T}
}

@article{vetter1986,
  title={Evaluation of a prototype dual-energy computed tomographic apparatus. II. Determination of vertebral bone mineral content},
  author={Vetter, JR and Perman, WH and Kalender, Willi A and Mazess, RB and Holden, JE},
  journal={Medical physics},
  volume={13},
  number={3},
  pages={340--343},
  year={1986},
  publisher={Wiley Online Library}
}

@Article{wux2017,
  author    = {Xiaomei Wu and Qian Wang and Jinlei Ma and Wei Zhang and Po Li and Zheng Fang},
  journal   = {Review of Scientific Instruments},
  title     = {A hyperspectral {X}-ray computed tomography system for enhanced material identification},
  year      = {2017},
  month     = aug,
  number    = {8},
  pages     = {083111},
  volume    = {88},
  doi       = {10.1063/1.4998991},
  publisher = {{AIP} Publishing},
}

@article{xuey2019,
  title={Accurate multi-material decomposition in dual-energy CT: A phantom study},
  author={Xue, Yi and Jiang, Yangkang and Yang, Chunlin and Lyu, Qihui and Wang, Jing and Luo, Chen and Zhang, Luhan and Desrosiers, Catherine and Feng, Kun and Sun, Xiaonan and others},
  journal={IEEE Transactions on Computational Imaging},
  volume={5},
  number={4},
  pages={515--529},
  year={2019},
  publisher={IEEE}
}

@article{yingz2006,
  title={Dual energy computed tomography for explosive detection},
  author={Ying, Zhengrong and Naidu, Ram and Crawford, Carl R},
  journal={Journal of X-ray Science and Technology},
  volume={14},
  number={4},
  pages={235--256},
  year={2006},
  publisher={IOS Press}
}

@article{korobkin2024isotopic,
  title={Isotopic gamma lines for identification of shielding materials},
  author={Korobkin, Oleg and Klasky, Marc L and Khatiwada, Ajeeta and McCann, Michael},
  journal={Applied Radiation and Isotopes},
  volume={212},
  pages={111422},
  year={2024},
  publisher={Elsevier}
}

@article{ensslin1991principles,
  title={Principles of neutron coincidence counting},
  author={Ensslin, N and others},
  journal={Passive Nondestructive Assay of Nuclear Materials},
  volume={550},
  pages={457--492},
  year={1991},
  publisher={US Nuclear Regulatory Commission, NUREG/CR-5550}
}

@article{hage1985factorial,
  title={On the factorial moments of the neutron multiplicity distribution of fission cascades},
  author={Hage, W and Cifarelli, Donato Michele},
  journal={Nuclear Instruments and Methods in Physics Research Section A: Accelerators, Spectrometers, Detectors and Associated Equipment},
  volume={236},
  number={1},
  pages={165--177},
  year={1985},
  publisher={Elsevier}
}

@article{dubi2018mass,
  title={Mass uncertainty in neutron multiplicity counting associated with the uncertainty on the fission multiplicity factorial moments},
  author={Dubi, C and Heger, G and Ocherashvili, A and Pedersen, B},
  journal={Nuclear Instruments and Methods in Physics Research Section A: Accelerators, Spectrometers, Detectors and Associated Equipment},
  volume={902},
  pages={83--87},
  year={2018},
  publisher={Elsevier}
}

@article{cifarelli1986models,
  title={Models for a three-parameter analysis of neutron signal correlation measurements for fissile material assay},
  author={Cifarelli, Donato Michele and Hage, W},
  journal={Nuclear Instruments and Methods in Physics Research Section A: Accelerators, Spectrometers, Detectors and Associated Equipment},
  volume={251},
  number={3},
  pages={550--563},
  year={1986},
  publisher={Elsevier}
}

@article{tang2024machine,
  title={A machine learning decision criterion for reducing scan time for hyperspectral neutron computed tomography systems},
  author={Tang, Shimin and Venkatakrishnan, Singanallur V and Chowdhury, Mohammad SN and Yang, Diyu and Gober, Megan and Nelson, George J and Cekanova, Maria and Biris, Alexandru S and Buzzard, Gregery T and Bouman, Charles A and others},
  journal={Scientific Reports},
  volume={14},
  number={1},
  pages={15171},
  year={2024},
  publisher={Nature Publishing Group UK London}
}

@inproceedings{chowdhury2023autonomous,
  title={Autonomous polycrystalline material decomposition for hyperspectral neutron tomography},
  author={Chowdhury, Mohammad Samin Nur and Yang, Diyu and Tang, Shimin and Venkatakrishnan, Singanallur V and Bilheux, Hassina Z and Buzzard, Gregery T and Bouman, Charles A},
  booktitle={2023 IEEE International Conference on Image Processing (ICIP)},
  pages={1280--1284},
  year={2023},
  organization={IEEE}
}

@article{xu2025swap,
  title={Swap-Net: A Memory-Efficient 2.5 {D} Network for Sparse-View {3D} Cone Beam {CT} Reconstruction to ICF Applications},
  author={Xu, Xiaojian and Klasky, Marc and McCann, Michael T and Hu, Jason and Fessler, Jeffrey A},
  journal={IEEE Transactions on Computational Imaging},
  year={2025},
  publisher={IEEE}
}

@article{lahiri2023sparse,
  title={Sparse-view cone beam {CT} reconstruction using data-consistent supervised and adversarial learning from scarce training data},
  author={Lahiri, Anish and Maliakal, Gabriel and Klasky, Marc L and Fessler, Jeffrey A and Ravishankar, Saiprasad},
  journal={IEEE Transactions on Computational Imaging},
  volume={9},
  pages={13--28},
  year={2023},
  publisher={IEEE}
}

@techreport{mattingly2009polyethylene,
  title={Polyethylene-reflected plutonium metal sphere: subcritical neutron and gamma measurements.},
  author={Mattingly, John K},
  year={2009},
  institution={Sandia National Laboratories (SNL), Albuquerque, NM, and Livermore, CA~…}
}

@techreport{langner1998application,
  title={Application guide to neutron multiplicity counting},
  author={Langner, DG and Stewart, JE and Pickrell, MM and Krick, MS and Ensslin, N and Harker, WC},
  year={1998},
  institution={Los Alamos National Laboratory, Los Alamos, NM}
}

@article{dubi2017estimating,
  title={Estimating the mass variance in neutron multiplicity counting—A comparison of approaches},
  author={Dubi, C and Croft, Stephen and Favalli, Andrea and Ocherashvili, A and Pedersen, B},
  journal={Nuclear Instruments and Methods in Physics Research Section A: Accelerators, Spectrometers, Detectors and Associated Equipment},
  volume={875},
  pages={125--131},
  year={2017},
  publisher={Elsevier}
}

@techreport{mitchell2014gadras,
  title={GADRAS Detector Response Function.},
  author={Mitchell, Dean J and Harding, Lee and Thoreson, Gregory G and Horne, Steven M},
  year={2014},
  institution={Sandia National Lab.(SNL-NM), Albuquerque, NM (United States)}
}

@conference{LA-UR-24-22644,
title = {Multimodal Material Identification Reconstruction Algorithms},
author = {Marc Louis Klasky and Oleg Korobkin and Michael R. James and Mark Andrew Nelson},
number = {LA-UR-24-22644},
note = {Sponsor: USDOE National Nuclear Security Administration (NNSA). Office of Defense Nuclear Nonproliferation R\&D (NA-22)},
year = {2024},
note = {Sponsor Update ; 2024-03-07 - 2024-03-07 ; Los Alamos, New Mexico, United States},
language = {English},
howpublished = {Sponsor Update ; 2024-03-07 - 2024-03-07 ; Los Alamos, New Mexico, United States},
note = {Other Numbers: LA-UR-24-22644},
note = {Version: 1},
url = {https://laauthors.lanl.gov/record/file?id=perma%3Alanl%2Frl-dv-0000425578%2F3},
note = {Repository ID: perma:lanl/rl-dv-0000425578},
note = {Dataset: rassti},
}

@conference{LA-UR-23-23930,
title = {Multimodal Radiographic Reconstruction Algorithms: Combining Radiography and Spectrometry for Material Identification},
author = {Oleg Korobkin and Marc Louis Klasky and Michael Thompson McCann and Jason Ray Matheny},
number = {LA-UR-23-23930},
note = {Sponsor: USDOE National Nuclear Security Administration (NNSA). Office of Defense Nuclear Nonproliferation R\&D (NA-22)},
year = {2023},
note = {NSARD 2023 Programmatic Review Meeting ; 2023-04-18 - 2023-04-21 ; Richland, Washington, United States},
language = {English},
howpublished = {NSARD 2023 Programmatic Review Meeting ; 2023-04-18 - 2023-04-21 ; Richland, Washington, United States},
note = {Other Numbers: LA-UR-23-23930},
note = {Version: 1},
url = {https://laauthors.lanl.gov/record/file?id=perma%3Alanl%2Frl-dv-0000404992%2F3},
note = {Repository ID: perma:lanl/rl-dv-0000404992},
note = {Dataset: rassti},
}

@conference{LA-UR-24-22707,
title = {Multimodal Material Identification Reconstruction Algorithms and Surrogate Modeling},
author = {Marc Louis Klasky and Oleg Korobkin and Balasubramanya T. Nadiga and Michael R. James and Mark Andrew Nelson},
number = {LA-UR-24-22707},
note = {Sponsor: USDOE National Nuclear Security Administration (NNSA). Office of Defense Nuclear Nonproliferation R\&D (NA-22)},
year = {2024},
note = {NSARD ; 2024-04-17 - 2024-04-17 ; Lemont, Illinois, United States},
language = {English},
howpublished = {NSARD ; 2024-04-17 - 2024-04-17 ; Lemont, Illinois, United States},
note = {Other Numbers: LA-UR-24-22707},
note = {Version: 1},
url = {https://laauthors.lanl.gov/record/file?id=perma%3Alanl%2Frl-dv-0000423335%2F3},
note = {Repository ID: perma:lanl/rl-dv-0000423335},
note = {Dataset: rassti},
}

@conference{LAPR-2021-037525,
title = {Comparison of the {F}eynman-{Y} and the {R}ossi-alpha methods for subcritical systems},
author = {Mark Nelson and Alex Mcspaden and George McKenzie and Jesson Hutchinson and Travis Baugher},
number = {LAPR-2021-037525},
journal = {International Conference on Physics of Reactors, PHYSOR 2018: Reactor Physics Paving the Way Towards More Efficient Systems},
volume = {Part F168384-5},
pages = {3103-3114},
year = {2018},
note = {Published as: International Conference on Physics of Reactors, PHYSOR 2018: Reactor Physics Paving the Way Towards More Efficient Systems ; Part F168384-5 ; 3103-3114 ; 2018},
note = {2018 International Conference on Physics of Reactors: Reactor Physics Paving the Way Towards More Efficient Systems, PHYSOR 2018 ; 04/22/2018 - 04/26/2018 ; Cancun, , MEX},
howpublished = {2018 International Conference on Physics of Reactors: Reactor Physics Paving the Way Towards More Efficient Systems, PHYSOR 2018 ; 04/22/2018 - 04/26/2018 ; Cancun, , MEX},
isbn = {9781713808510},
note = {Other Numbers: LAPR-2021-037525},
note = {Repository ID: perma:lanl/rl-dv-0000439633},
note = {Dataset: lapr},
}

@article{khatiwada2023machine,
  title={Machine Learning technique for isotopic determination of radioisotopes using HPGe $\gamma$-ray spectra},
  author={Khatiwada, Ajeeta and Klasky, Marc and Lombardi, Marcie and Matheny, Jason and Mohan, Arvind},
  journal={Nuclear Instruments and Methods in Physics Research Section A: Accelerators, Spectrometers, Detectors and Associated Equipment},
  volume={1054},
  pages={168409},
  year={2023},
  publisher={Elsevier}
}

@article{bell2025learning,
  title={Learning robust parameter inference and density reconstruction in flyer plate impact experiments},
  author={Bell, Evan and Serino, Daniel A and Southworth, Ben S and Wilcox, Trevor and Klasky, Marc L},
  journal={Scientific Reports},
  volume={15},
  number={1},
  pages={40846},
  year={2025},
  publisher={Nature Publishing Group UK London}
}

\end{document}